\documentclass[onecolumn,preprint,superscriptaddress]{revtex4-2}
\usepackage{bm}
\usepackage[colorlinks=true,linkcolor=blue,citecolor=blue]{hyperref}
\usepackage{times}
\usepackage{amsmath}
\usepackage{amssymb}
\usepackage{amsthm}
\usepackage{amsfonts}
\usepackage{enumerate}
\usepackage{latexsym}
\usepackage{ifpdf}
\usepackage{natbib}
\usepackage{psfrag}
\newcommand{\beq}{\begin{equation}}
	\newcommand{\eeq}{\end{equation}}
\usepackage{graphicx}
\usepackage{makeidx}
\usepackage{xcolor}
\usepackage{array}
\usepackage{multirow}
\usepackage{siunitx}
\usepackage{}
\usepackage{physics}
\usepackage{soul}
\def \w{\omega}
\def \cd{c^{\dagger}}
\def \cb{\bar{c}}

\def \fb{\bar{f}}
\def \t{\tau}
\def \beq{\begin{equation}}
	\def \beqn{\begin{align}}
		\def \eeq{\end{equation}}
	\def \eeqn{\begin{align}}
		\def \deltau{\partial_{\tau}}

\begin{document}
	
	\title{Quantum fluctuations lead to glassy electron dynamics in the good metal regime of electron doped KTaO$_3$}
	
	\author {Shashank Kumar Ojha}
	\email{shashank@iisc.ac.in}
	\affiliation{Department of Physics, Indian Institute of Science, Bengaluru 560012, India}
	\author {Sankalpa Hazra}
	\altaffiliation{Contributed equally}
	\affiliation{Department of Physics, Indian Institute of Science, Bengaluru 560012, India}
	\affiliation{Department of Materials Science and Engineering, The Pennsylvania State University, University Park, PA 16802, USA}
	\author {Surajit Bera}
	\altaffiliation{Contributed equally}
	\affiliation{Department of Physics, Indian Institute of Science, Bengaluru 560012, India}
	\author {Sanat Kumar Gogoi}
	\affiliation{Department of Physics, Indian Institute of Science, Bengaluru 560012, India}
	\affiliation{Department of Physics, Digboi College, Digboi 786171, India}
	\author {Prithwijit Mandal}
	\affiliation{Department of Physics, Indian Institute of Science, Bengaluru 560012, India}
	\author {Jyotirmay Maity}
	\affiliation{Department of Physics, Indian Institute of Science, Bengaluru 560012, India}
	\author {A. Gloskovskii}
	\affiliation{Deutsches Elektronen-Synchrotron DESY, 22607 Hamburg, Germany}
	\author {C. Schlueter}
	\affiliation{Deutsches Elektronen-Synchrotron DESY, 22607 Hamburg, Germany}
	\author {Smarajit Karmakar}
	\affiliation{ Tata Institute of Fundamental Research, 36/P, Gopanpally Village, Serilingampally Mandal, Ranga Reddy District, Hyderabad, 500107, India}
	\author {Manish Jain}
	\affiliation{Department of Physics, Indian Institute of Science, Bengaluru 560012, India}
	\author {Sumilan Banerjee}
	\email{sumilan@iisc.ac.in}
	\affiliation{Department of Physics, Indian Institute of Science, Bengaluru 560012, India}
	\author {Venkatraman Gopalan}
	\affiliation{Department of Materials Science and Engineering, The Pennsylvania State University, University Park, PA 16802, USA}
	\author {Srimanta Middey}
	\email{smiddey@iisc.ac.in}
	\affiliation{Department of Physics, Indian Institute of Science, Bengaluru 560012, India}

	\begin{abstract}

		\textbf{Abstract: One of the central challenges in condensed matter physics is to comprehend systems that have strong disorder and strong interactions. In the strongly localized regime, their subtle competition leads to glassy electron dynamics which ceases to exist well before the insulator-to-metal transition is approached as a function of doping. Here, we report on the discovery of glassy electron dynamics deep inside the good metal regime of an electron-doped quantum paraelectric system: KTaO$_{3}$. We reveal that upon  excitation of electrons from defect states to the conduction band, the excess  injected carriers in the conduction band relax in a stretched exponential manner with a large relaxation time, and the system evinces simple aging phenomena - a telltale sign of glassy dynamics. Most significantly, we observe a critical slowing down of carrier dynamics below 35 K, concomitant with the onset of quantum paraelectricity in the undoped KTaO$_3$. Our combined investigation using second harmonic generation technique, density functional theory and phenomenological modeling demonstrates quantum fluctuation-stabilized soft polar modes as the impetus for the glassy behavior. This study addresses one of the most fundamental questions regarding the potential promotion of glassiness by quantum fluctuations and opens a route for exploring  glassy dynamics of electrons in a well-delocalized regime.}

	\end{abstract}
	
	\maketitle
	
\begin{figure*}[htp]
		\centering{
			{~}\hspace*{-0.2cm}
			\includegraphics[scale=.30]{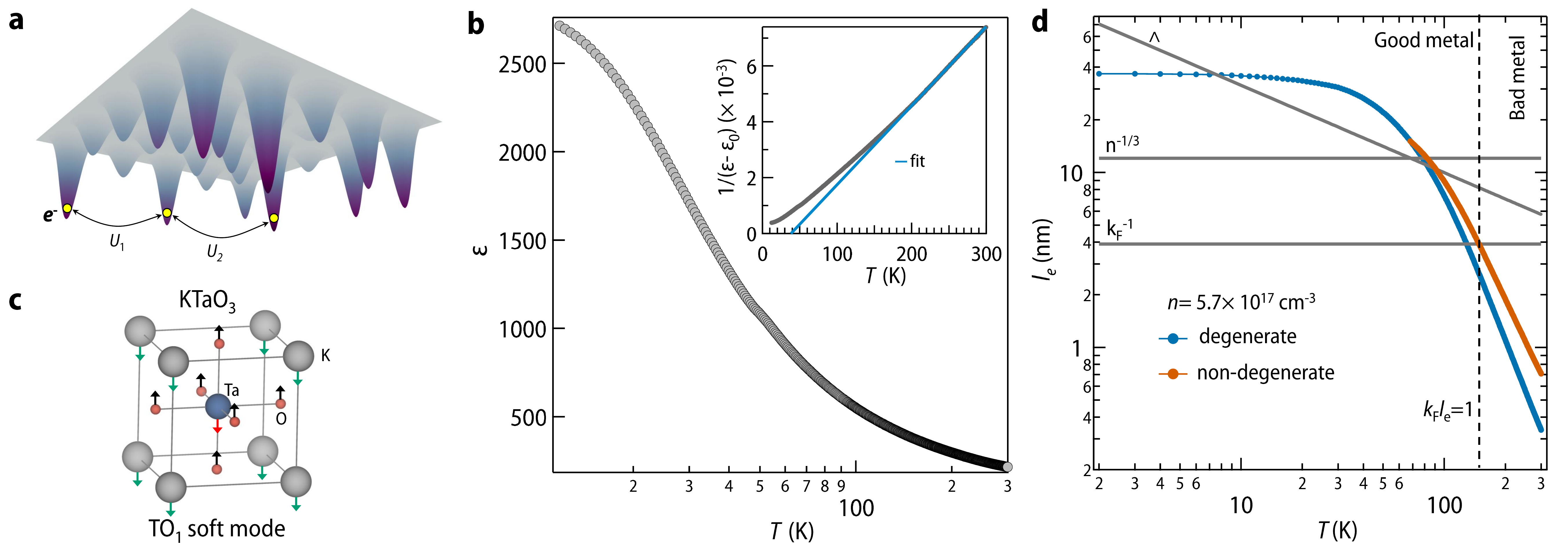}
			\caption{\textbf{Electron glass, quantum paraelectricity and temperature dependence of electron's mean free path in KTaO$_{3-\delta}$}  \textbf{a.} A schematic to describe the concept of electron glass in a disordered insulator with highly localized electronic states. Here the solid colored cones represent the random disorder potential and yellow-filled circles represent electrons trapped in them. Disorder tries to make the distribution of doped carriers random, however long-range Coulomb interactions ($U_1$, $U_2$) try to make the distribution homogeneous leading to electronic frustration. \textbf{b.} Temperature dependent dielectric constant ($\epsilon$)  of pristine (001) oriented KTaO$_3$ single crystal recorded at an AC frequency of 10 kHz, taken from our previous paper~\cite{Ojha:2021p085120}. We further note that the value of $\varepsilon$ for our sample appears to be slightly lower than the reported values~\cite{Rowley:2014p367}. We attribute this difference to the difference in the sample preparation process, which may add slight oxygen vacancies, even in the pristine, as-received crystal~\cite{Salce1994}. Sitting on the verge of quantum critical point, KTaO$_3$ is long known for its peculiar dielectric properties~\cite{Lemanov:2007p97} wherein the onset of quantum fluctuation leads to a marked departure from classical paraelectric behavior below 35 K leading to saturation of $\varepsilon$ at low temperature. This is further evident from the modified Curie–Weiss fitting ($\varepsilon$ = $\varepsilon_0$+	$\frac{C_W}{T-\theta_{CW}}$, $\varepsilon_0$ is temperature independent component, $C_W$ is Curie-Weiss constant and $\theta_{CW}$ is Curie-Weiss temperature, respectively) shown in the inset of panel b. $\theta_{CW}$ obtained from the fitting is found to be 37 K.  \textbf{c.} Unit cell of pristine KTaO$_3$ along with the transverse optical soft mode. \textbf{d.} Temperature-dependent electron's mean free path $l_e$ for oxygen-deficient KTO calculated within
				the Drude-Boltzmann picture. Blue and orange curves correspond to degenerate and non-degenerate cases respectively. $\Lambda$ denotes the thermal de Broglie wavelength. A dotted vertical line marks the temperature above which $l_e$ becomes shorter than the inverse of Fermi wave-vector ($k_F$) and the sample
				crosses over from good metal to a bad metal phase. For details of the calculations, we refer to Supplementary Note 1 which significantly overlaps with our earlier work~\cite{Ojha:2020p2000021}. Source data are provided as a Source Data file.}\label{fig:1}}
	\end{figure*}

	\textbf{Introduction}: The notion of glassy dynamics associated with the electronic degree of freedom in condensed matter systems was first envisaged by Davies, Lee and Rice in 1982~\cite{Davies:1982p758}. Building heavily on Anderson's seminal work on the localization of wave functions in random disordered  media~\cite{Anderson:1958p1492}, it was predicted that, in a disordered insulator with highly localized electronic states, the interplay between disorder and long-range Coulomb interaction (Fig. \ref{fig:1}a) should precipitate electronic frustration in real space. Such a scenario would result in a rugged energy landscape with numerous metastable states, leading to the emergence of electron glass~\cite{dobrosavljevic:2012p,pollak:2013p,Amir:2011p235}. This hypothesis was soon tested on various strongly localized electronic systems, including granular metals, crystalline and amorphous oxides, and later, it was also extended to doped semiconductors and two-dimensional electron gases~\cite{Kar:2003p216603,dobrosavljevic:2012p,pollak:2013p,Amir:2011p235,Mahmood:2021p627}. The typical relaxation time in these systems ranges from a few seconds to several hours, making them an obvious choice for studying glassy physics in laboratory timescales. What makes these systems even more intriguing is the array of perturbations that one can use to effectively drive them away from equilibrium~\cite{pollak:2013p}. Furthermore, due to the light mass of electrons, electron glasses are highly susceptible to quantum fluctuations. This aspect introduces additional complexities in understanding the behavior of electron glasses~\cite{Pastor:1999p4642}.

	In the antithetical regime of highly delocalized electrons i.e. in a metal, the screening effect significantly reduces the strength of the electron-electron and electron-impurity interactions. Consequently, such system generally possesses a non-degenerate ground state with a well-defined Fermi surface. As a result, the dynamics of glassy behavior, which involves the existence of multiple, competing ground states, is incompatible with the behavior of metals. In fact, the manifestation of glassiness fades away considerably prior to the transition from insulator to metal, and there is an absolute lack of any substantiated indication of the presence of glassiness within a good metal regime~\cite{dobrosavljevic:2012p} where quasi-particle mean free path is larger than the electron's wavelength.

	In this work, we report on the discovery of glassy dynamics of conduction electrons in an electron-doped quantum paraelectric system, namely KTaO$_3$, in a good metal regime. Even more surprising observation is that glassiness is found to appear in a regime where quantum fluctuations are inherently present in the system.  In pristine KTaO$_3$, the quantum fluctuations associated with the zero point motion of the atoms prohibits the onset of ferroelectric order below 35 K (Fig. \ref{fig:1}b), and the system's properties are typically governed by the presence of an associated low energy transverse optical phonon (Fig. \ref{fig:1}c) popularly known as soft polar mode~\cite{Chandra:2017p112502}. Our combined transport and optical second harmonic generation measurements find that properties associated with the soft polar mode are preserved even deep inside the metallic regime. Most importantly, such soft modes are found to be directly responsible for emergent glassy dynamics at low temperatures, which is further corroborated by our theoretical calculations. Our observation is one of the rarest examples where quantum fluctuation, which is generally considered as a bottleneck for electron glass formation~\cite{Pastor:1999p4642}, is ultimately accountable for the appearance of glassy dynamics in a good metallic phase. 
	
	\noindent \textbf{Results}:

	\textbf{Demonstration of good metallic behavior}: Due to the remarkable applications of KTO in the field of spintronics and prospects of studying emergent physics close to the ferroelectric quantum critical point, several successful attempts have been made in recent times to introduce free carriers in the bulk as well as at the surface or interface of KTO and
	a wide range of phenomena ranging from topology to 2D superconductivity have been reported ~\cite{Gupta:2022p2106481,Changjiang:2021p716,Ren:2022peabn4273,Ojha:2023p126}. However, so far there has been no report about glassy dynamics in electron doped KTO. For the current investigation, metallic samples have been prepared by introducing oxygen vacancies in pristine single crystalline (001) oriented KTaO$_3$ substrate (for details see methods section and references ~\cite{Ojha:2020p2000021,Ojha:2021p085120}). These samples were found to exhibit quantum oscillations below 10 K~\cite{Ojha:2020p2000021}, which is signature of a good metal with well-defined Fermi surface. To further testify this, we have also computed temperature-dependent mean free path of electrons ($l_e$) within Drude–Boltzmann picture. Fig. \ref{fig:1}d shows the corresponding plots for degenerate and non-degenerate case. A dotted vertical line marks the temperature above which $l_e$ becomes shorter than the inverse of Fermi wave-vector ($k_F$) and the sample crosses over from good metal to a bad metal phase~\cite{Lin:2017p41}. In the current work, observation of glassy dynamics is inherently constrained to temperatures which is much lower than crossover temperature to bad metal phase and hence for all practical purposes our electron doped KTaO$_3$ system can be considered as a good metal with well-defined scattering.
	
\begin{figure*}[ht]
		\centering{
			{~}\hspace*{-0.3cm}
			\includegraphics[scale=.35]{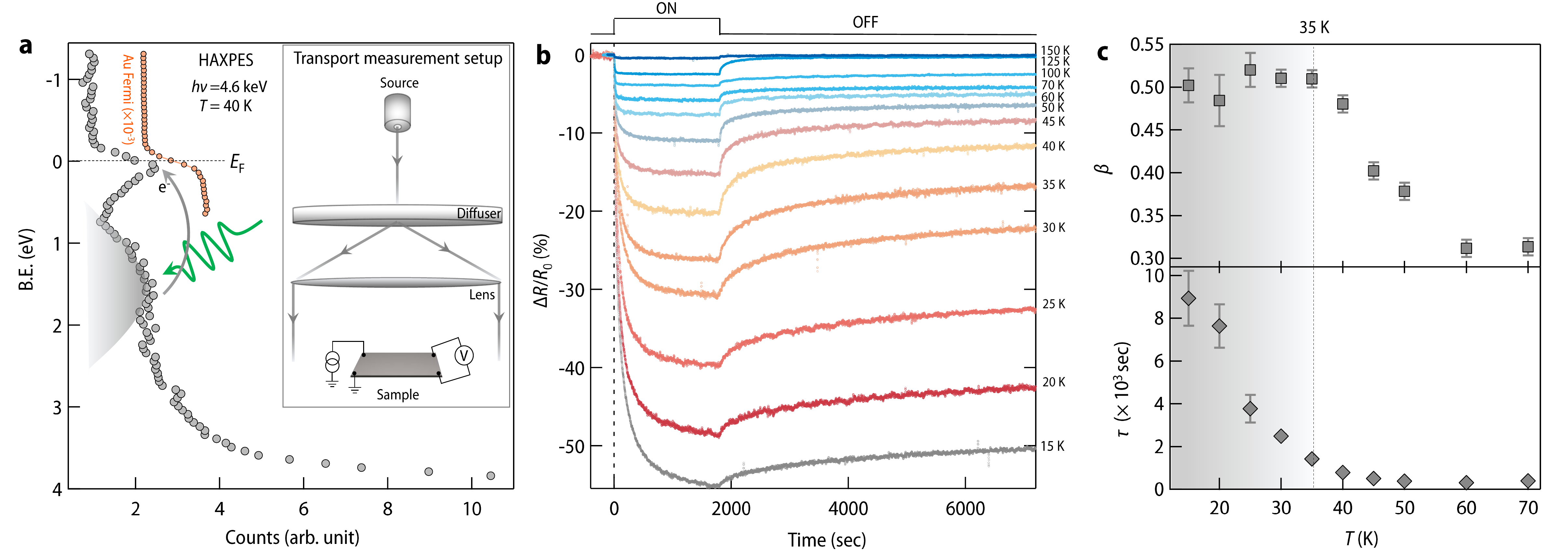}
			\caption{\textbf{HAXPES and transport measurements under light}  \textbf{a.} Near Fermi level electronic states measured at 40 K using hard X-ray photoelectron spectroscopy on the KTaO$_3$$_{-\delta}$ sample. Binding energy was corrected by taking Au Fermi level as a reference. In the figure, Au Fermi level intensity has been reduced by thousand times and shifted rightward for comparison.  Inset shows the setup for transport measurement under light illumination (see methods for more details). Green arrow shows the incoming photon which excites trapped electrons to the conduction band (shown with a slate-gray color). A slate-gray shade has been used to highlight the presence of defect states. \textbf{b.} Temporal evolution of resistance under green light illumination ($\lambda$ = 527 nm, power = 145 $\mu$ watt) for 30 minutes measured at several fixed temperatures. After 30 minutes, resistance relaxation was observed in dark for the next 1.5 hours. For comparative analysis, change in resistance has been converted into relative percentage change ($\Delta$$R$/$R_0$)$\cross$100. Additional measurements with red light have been shown in Supplementary Note 5 and Supplementary Figure 5. \textbf{c.} Temperature dependence of  stretching exponent ($\beta$) and relaxation time ($\tau$) obtained from fitting of resistance relaxation in light off stage with a stretched exponential function. A slate-gray shade has been used to highlight the distinct behavior of $\beta$ and $\tau$ below 35 K. The error bar in $\beta$ and $\tau$ has been estimated from the variation of corresponding parameters which results in similar fitting. We further emphasize that, while the value $\beta$ has been obtained by fitting the data for 1.5 hours, the same value is obtained even when the data is fitted for a longer period of up to 24 hours Supplementary Note 6 and Supplementary Figure 6). Source data are provided as a Source Data file.}   \label{fig:2}}
	\end{figure*}

	\textbf{Demonstration of glassy dynamics}: Oxygen vacancy creation in KTaO$_3$ not only adds free electrons  but also leads to the formation of highly localized defect states. To determine the exact position of the defect states, valence band spectrum has been mapped out by using hard X-ray photoelectron spectroscopy (HAXPES) at P22 beamline of PETRA III, DESY (see methods for more details). Apart from the well-defined quasi-particle peak,  mid-gap states centered at 1.6-1.8 eV is observed (Fig. \ref{fig:2}a), which arises due to the clustering of oxygen vacancies~\cite{Ojha:2021p085120}.

	In this work, we utilize these defect states to perturb the system by selective excitation of trapped electrons  to the conduction band via sub-bandgap light illumination. Subsequently, the system's response is studied by monitoring the temporal evolution of the electrical resistance at a fixed temperature (see inset of Fig. \ref{fig:2}a for transport measurement set-up). For each measurement, the sample was first cooled down to the desired temperature. Once the temperature stabilizes, the system was driven out of equilibrium by shining light for half an hour, thereafter resistance relaxation was observed for the next 1.5 hours in dark condition. For the next measurement, the sample was heated to room temperature where the original resistance is recovered. Fig. \ref{fig:2}b shows one set of data recorded with green light ($\lambda$=527 nm) at several fixed temperatures ranging from 15 K to 150 K.

	At first glance, Fig. \ref{fig:2}b reveals a striking temperature dependence of the photo-doping effect (also see Supplementary Note 2 and Supplementary Figure 2) and the way the system relaxes after turning off the light is also found to be strongly temperature dependent. Further analysis reveals that in the off-stage, the resistance relaxes in a stretched exponential manner [exp(-($t$/$\tau$)$^\beta$) where $\tau$ is the relaxation time and $\beta$ (stretching exponent) $<$ 1 (see Supplementary Figure 3 for fitting of few representative data)].  Such stretched exponential relaxation is very often considered as a signature of glassy dynamics and has been observed in variety of glassy systems~\cite{pollak:2013p,Du:2007p111,VidalRussell:2000p695}. In glass physics, it is commonly accepted that such stretching is due to a distribution of relaxation times arising from the disorder-induced heterogeneity~\cite{Sillescu:1999p81, Ediger:2000p99}. In the present case, the distribution of relaxation times would correspond to multiple relaxation channels for the electron-hole recombination~\cite{Johnston:2006p184430}. The microscopic origin behind the spatial separation of electron-hole pair and resultant non-exponential relaxation will be discussed later. The temperature evolution of $\beta$ and $\tau$ obtained from the fitting further reveals a substantial increase in relaxation time  below 50 K with a power law behavior ($\tau\sim T^{-2.8}$, see Supplementary Note 4 and Supplementary Figure 4) followed by constant $\beta \approx$ 0.5 below 35 K (Fig. \ref{fig:2}c). This observation is remarkable given the fact that this crossover temperature roughly coincides with the onset of quantum fluctuation in undoped  KTaO$_3$.
	
\begin{figure*}[htp]
		\centering{
			{~}\hspace*{-0.40cm}
			\includegraphics[scale=.36 ]{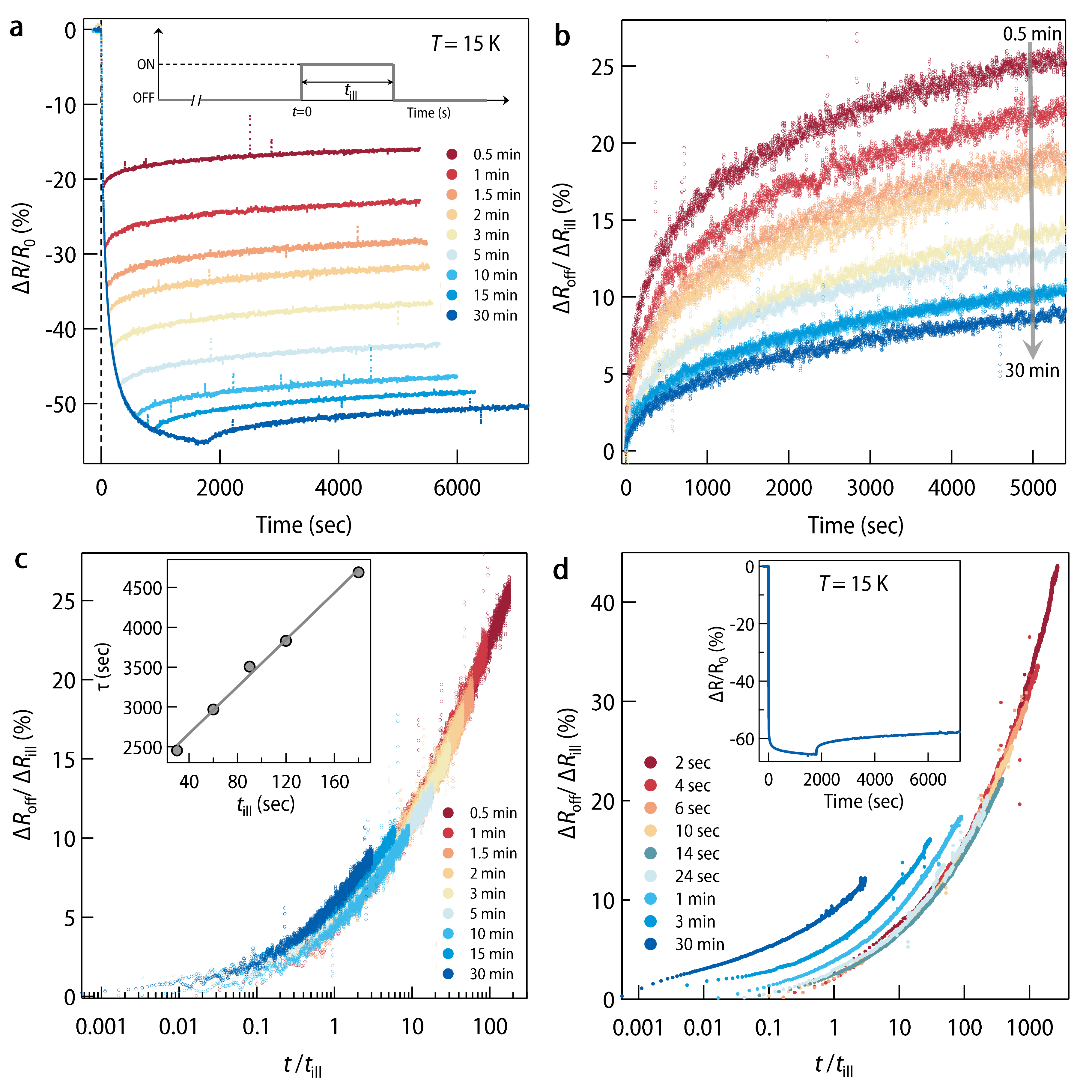}
			\caption{\textbf{Aging phenomena}  \textbf{a.} Relative percentage change in resistance ($\Delta$$R$/$R_0$)$\cross$100 measured at 15 K for different $t$$_\text{ill}$ ranging from 0.5 min to 30 min. After turning off the light, resistance relaxation was observed for the next 1.5 hours in every case. \textbf{b.} Temporal evolution of resistance relaxation in off-stage for different values of $t$$_\text{ill}$. Since the amount of doping depends on the duration of light illumination, the change in resistance in the off-stage ($\Delta$$R$$_\text{off}$) has been normalized with  a total drop in resistance ($\Delta$$R$$_\text{ill}$) at the end of illumination~\cite{Lee:2005L439}. \textbf{c.} Upon re-scaling the time axis with $t$$_\text{ill}$, all the curves in off-stage (except $t$$_\text{ill}$ = 30 min) are found to fall on a universal curve. Inset shows the linear relationship between $\tau$ and $t$$_\text{ill}$ which is the defining criteria for simple/full aging. \textbf{d.} Simple aging observed for another sample at 15 K. Inset shows one representative resistance relaxation data at 15 K. The sheet resistance of this sample is approximately 22 k$\Omega$/sq. (at room temperature) which is much larger than the one discussed throughout this manuscript which has around 200 $\Omega$/sq. emphasizing that the oxygen vacancy concentration for this sample is much lower. Source data are provided as a Source Data file.} \label{fig:3}}
	\end{figure*}

	One critical test to confirm  glassiness is the observation of  the aging phenomenon wherein the system's response depends on its age~\cite{Amir:2012p1850}. More precisely, older systems are found to relax more slowly than younger ones. To examine this, we prepared the system of desired age by tuning the duration of light illumination ($t$$_\text{ill}$), and measurements similar to that shown in Fig. \ref{fig:2}b were  performed at 15 K (Fig. \ref{fig:3}a).  For further analysis, we only focus on relaxation after turning off the light. In Fig. \ref{fig:3}b we plot the change in resistance in the off-stage ($\Delta$$R$$_\text{off}$) normalized with a total drop in resistance ($\Delta$$R$$_\text{ill}$) at the end of illumination. Evidently, with increasing $t$$_\text{ill}$ system's response becomes more and more sluggish. More precisely, $\tau$ obtained from the fitting is found to scale linearly with $t$$_\text{ill}$ (inset of Fig. \ref{fig:3}c). This is the defining criteria for simple or full aging~\cite{Amir:2012p1850} which is more clear in Fig. \ref{fig:3}c where all the curves can be collapsed to a universal curve by normalizing the abscissa by $t$$_\text{ill}$. A careful look at Fig. \ref{fig:3}c reveals that for larger values of $t$$_\text{ill}$, curves start to deviate from the universal scaling. This is much clear in Fig. \ref{fig:3}d which contains a similar set of scaling data for another sample with a lower carrier concentration. Such an observation is consistent with the criteria for aging that $t$$_\text{ill}$ should be much less than the time required to reach the new equilibrium under perturbation. As evident from Fig. \ref{fig:3}a and inset of Fig. \ref{fig:3}d, above a critical value of $t$$_\text{ill}$ there is little change in resistance upon shining the light any further. This signifies that the system is closer to its new equilibrium and hence aging ceases to hold at a higher $t$$_\text{ill}$.

	\textbf{Presence of polar nano regions \& importance of quantum fluctuations}:
	As mentioned earlier, the glassy behavior of electrons in conventional electron glasses results from the competition between disorder and Coulomb interactions and is only applicable in strongly localized regime~\cite{dobrosavljevic:2012p,pollak:2013p,Amir:2011p235}. As glassiness is observed within a good metal regime ($k_F$$l_e$$>$1) in our oxygen-deficient KTaO$_3$ samples,  we need to find other processes responsible for the electron-hole separation and complex glassy relaxations in the present case. In the context of electron doping in another well-studied quantum paraelectric SrTiO$_3$, a large lattice relaxation (LLR) model involving deep trap levels~\cite{Lang:1977p635} has been associated as a dominant cause for prohibiting electron-hole recombination~\cite{Tarun:2013p187403,Kumar:2015p205117}. However, our analysis of $\tau$ vs. $T$ (Supplementary Note 7 and Supplementary Figure 7) does not support the applicability of the LLR model in the present case. 
	In sharp contrast, we will conclusively demonstrate here that the effective charge separation in such systems directly correlates with the appearance of polar nano regions (PNRs), which arise as a direct consequence of the defect dipoles present in a highly polarizable lattice of quantum paraelectric~\cite{Vugmeister:1990p993}.
	
\begin{figure*}[htp]
		\centering{
			{~}\hspace*{0cm}
			\includegraphics[scale=.36 ]{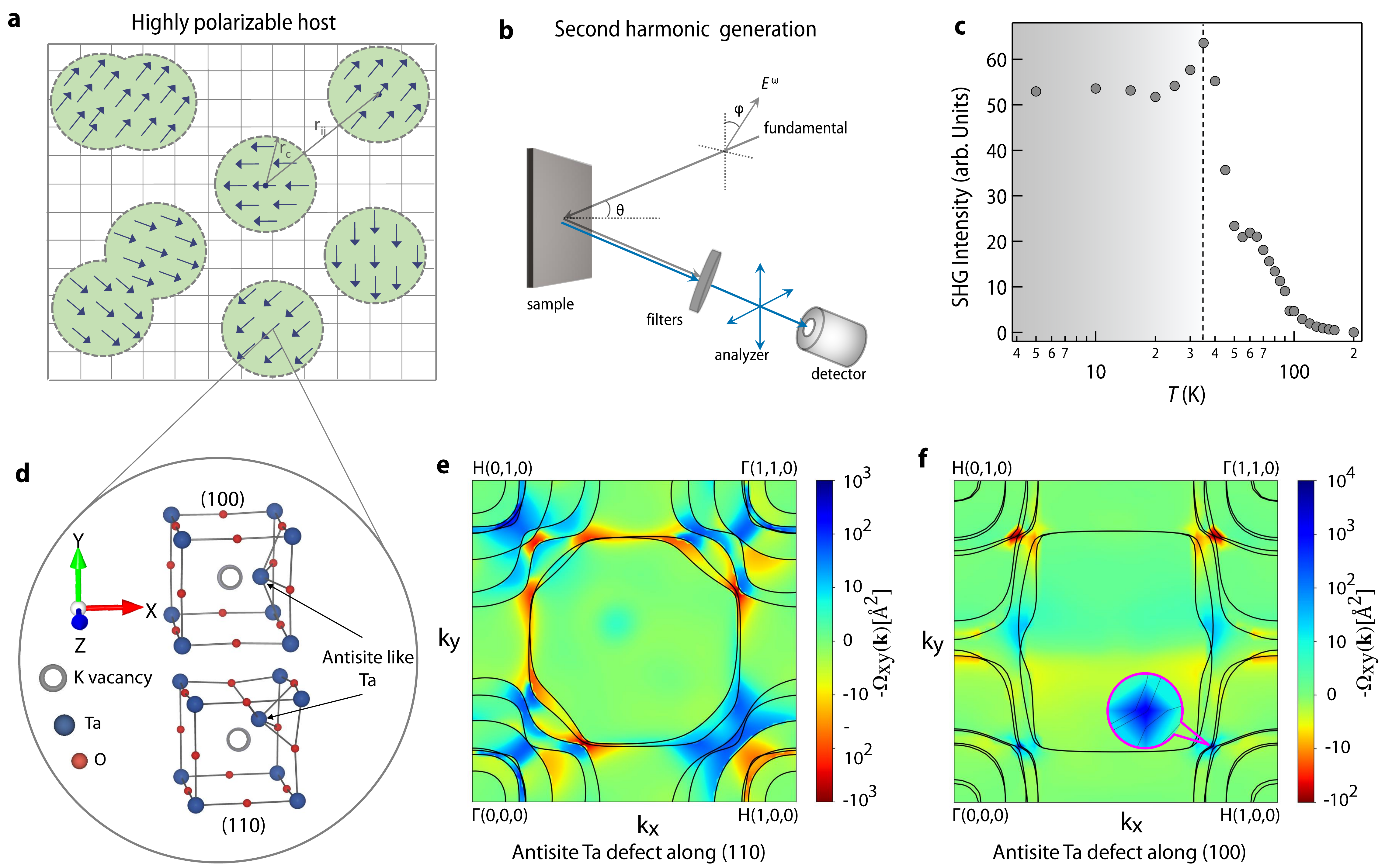}
			\caption{\textbf{Polar nano regions, second harmonic generation measurement and density functional theory} \textbf{a.} A schematic depicting real space random distribution of polar nano regions in a highly polarizable host lattice.  \textbf{b.} A schematic of the experimental setup for measuring optical second harmonic generation signal in reflection geometry (for more details see methods section). \textbf{c.} Temperature-dependent second harmonic generation intensity measured on KTaO$_3$$_{-\delta}$ sample.  A slate-gray shade has been used to highlight the distinct second harmonic generation signal below 35 K. \textbf{d.} A portion of the relaxed structure of supercell of size 2$\cross$2$\cross$2  of KTaO$_3$ with Ta off-centering along the (100) and (110) direction around K vacancy. For more details, see the methods section, Supplementary Note 12, and Supplementary Figure 12. \textbf{e.} Plot of Berry curvature $\Omega_{xy}(\vb{k})$ (shown in color-map) for $k_z =0$ and bands (solid contours) intersecting the Fermi surface  for the system having an antisite-like Ta defect along (110) direction. \textbf{f.} Same as \textbf{e.}  for the system with an antisite-like Ta defect along (100) direction. The inset shows the spin-orbit induced avoided crossing of bands yielding large contributions to the Berry curvature. We get average polarization of 5.8 $\mu C/cm^2$  and 11.1 $\mu C/cm^2$ for Ta antisite-like defect along (110) and  (100) respectively for the supercell of size $2\cross 2\cross 2$. We also note that,  we do not observe any polarization in pristine KTaO$_3$. Source data are provided as a Source Data file.} \label{fig:4}}
	\end{figure*}

	In an ordinary dielectric host, an electric dipole can polarize the lattice only in its immediate vicinity and hence the correlation length ($r_c$) is generally of the order of unit cell length which further remains independent of temperature~\cite{Samara:2003pR367}. However the situation is drastically different in highly polarizable hosts such as KTaO$_3$ where the magnitude of $r_c$ is controlled by the polarizability of the lattice which is inversely proportional to the soft mode frequency ($\omega$$_s$). Since $\omega$$_s$ decreases with decreasing temperature, $r_c$ becomes large at lower temperatures. As a result, PNRs spanning several unit cells are formed around the defect dipole which is randomly distributed in the lattice (Fig. \ref{fig:4}a).

	While PNRs are quite well established in the insulating regime (Supplementary Note 8 and Supplementary Figure 8), they are expected to vanish in the metals  due to screening effects from free electrons. Surprisingly, several recent experiments have reported that PNRs can exist even in the metallic regime~\cite{Bussmann:2020p,Lu:2018p491,Wang:2019p61,Rischau:2017p643,Salmani:2020p6542}.  Motivated by these results, we have carried out temperature-dependent optical second harmonic generation (SHG) measurement (Fig. \ref{fig:4}b) which is a powerful technique to probe PNRs~\cite{Jang:2010p197601}. Fig. \ref{fig:4}c shows the temperature evolution of SHG intensity which is directly proportional to the volume density of PNRs. As evident, no appreciable SHG signal is observed at room temperature, however,  a strong signal enhancement is observed below 150 K, signifying the appearance of PNR below 150 K in our metallic sample. Since  this onset temperature exactly coincides with the temperature below which an appreciable  photo-doping effect is observed in our transport measurements (Supplementary Note 9 and Supplementary Figure 9), we believe that the internal electric field generated around such PNRs is the major cause behind driving apart the photo-generated electron-hole pairs in real space.   Another notable observation is that the SHG intensity is independent of the temperature below 35 K. This immediately reminds of the regime of quantum fluctuation which enforces a constant value of $\omega_s$ below 35 K~\cite{Chandra:2017p112502}. Since $r_c \propto \omega_s^{-1}$~\cite{Chandra:2017p112502}, our SHG measurement conclusively establishes that quantum fluctuation stabilized soft polar mode is retained even in metallic KTaO$_3$~\cite{Bauerle:1980p335} (also see Supplementary Note 10).
	
	We have also conducted an investigation into the primary defect dipoles responsible for the creation of PNRs in our samples. Considering that our samples were prepared through high-temperature annealing within an evacuated sealed quartz tube~\cite{Ojha:2021p085120}, there is a possibility of K vacancies due to its high volatility. The presence of a certain degree of K vacancy has been indeed observed in our HAXPES  measurements (Supplementary Note 11 and Supplementary Figure 11). It is widely established that, in  $AB$O$_3$ systems, the off-centering of substitute $B$ atom ($B$ antisite-like defect) in the presence of $A$ atom vacancy leads to a macroscopic polarization even in a non-polar matrix ~\cite{Gopalan:2007p449,Lee:2015p1314,Jang:2010p197601,Choi:2009p185502,KonstantinPRB035301:2017}.  Our density functional theory calculations (details are in the method section) considering Ta antisite-like defect has found significant Ta off-centering along [100] and [110] in the presence of K vacancy  (Fig. \ref{fig:4}d). Further, we have also computed induced polarization in the system following the modern theory of polarization where the change in macroscopic electric polarization is represented by a Berry phase~\cite{VanderbiltPRB1651:1993, VanderbiltPRB195118:2006, YaoPRL037204:2004, RestaRMP899:1994, RestaENews18:1997} and the non-zero Berry curvature i.e., Berry phase per unit area, is taken as a signature of finite polarization in the material.  In Fig.\ref{fig:4} e, f, we have plotted the Berry curvature in the plane $k_z =0$ ($\Omega_{xy}(\vb{k})$) for the Ta off-centering along [110] and [100], respectively. From the figure, it is clear that the large contributions to the Berry curvature are due to the avoided crossings of bands at the Fermi surface (see inset of Fig. \ref{fig:4}f) which are induced by spin-orbit coupling.

	\textbf{Phenomenological model to understand glassy dynamics}:
	We now discuss a possible mechanism for the emergent glassy dynamics in a metal where conduction electrons coexist with PNRs. Since the glassiness in the present case is observed in the dynamical relaxation of excess injected carriers in the conduction band, it is necessary to have a thorough understanding of the relaxation processes happening against the backdrop of randomly oriented PNRs.  In indirect-band semiconductors like KTaO$_3$ which have strong electron-lattice and defect-lattice coupling, the relaxation should be predominantly nonradiative and manifest itself as the emission of several low-energy phonons~\cite{pelant2012luminescence}. Further, since inter-band electron-phonon matrix element due to acoustic phonons are negligible, the electron-hole recombination may primarily involve soft polar modes at low temperatures, although it is not the lowest energy phonon~\cite{Perry:1989p8666}. While such a multi-phonon inter-band transition could lead to multichannel relaxation with large time scales~\cite{pelant2012luminescence}, it can never give rise to collective glassy behavior.

	Instead, we suggest the following scenario for the observed glassiness. As was previously mentioned, there is clear evidence that the internal electric field around PNRs has a significant impact on electron-hole recombination in our sample. It has been demonstrated previously~\cite{Vugmeister:1990p993} that the random interactions between PNRs (in the limit of dilute defect dipoles) leads to a dipole glass at low temperatures in KTaO$_3$. Recently, long-lived glass-like relaxations in SHG and Kerr signals  were observed in pristine KTO at temperatures below 50 K and was attributed to dipolar correlations among PNRs, further highlighting the potential role of PNRs in influencing relaxation properties~\cite{Cheng:2023p126902, Li:2023p064306}. Electrons and holes being charged particles would immediately couple to the complex electric field from the dipole glass and hence there is a chance that the glassy background of PNRs can induce glassiness to the free carriers in the system. 
	
	In order to study such a possibility, we consider a theoretical  model with the Hamiltonian,
	\begin{align}
		H&=H_{el}+H_{gl}+H_{el-gl}, \label{eq:Hamiltonian}
	\end{align}
	where $H_{el}=-\sum_{ij}t_{ij}c_i^\dagger c_j-\varepsilon_0\sum_{\alpha}f_\alpha^\dagger f_\alpha$ ($\varepsilon_0>0$) describes the electronic part in terms of creation and annihilation operators $c_i^\dagger,c_i$ ($i=1,\cdots,N_c$) and $f_\alpha^\dagger,f_\alpha$ ($\alpha=1,\cdots,N_f$) of the $N_c$ conduction and $N_f$ impurity electronic states, respectively. For our calculations, we consider various lattices and corresponding hopping amplitudes $t_{ij}$ to describe several different energy dispersions for the conduction band, e.g., a band with semicircular DOS $g(\epsilon)=(1/2\pi)\sqrt{W^2-\omega^2}\theta(W-|\omega|)$ with band width $W$ [$\theta(x)$ is heaviside step function] (Fig. \ref{fig:5}(a)), and a flat band with width $W=0$, as discussed in the Supplementary Note 13. 

	We take a flat impurity band at energy $-\varepsilon_0$, separated by a gap $\Delta=\varepsilon_0-W/2$ from the conduction band minimum. We set the chemical potential $\mu=0$, at the centre of conduction band (Fig. \ref{fig:5}(a)). 
	
	To model the dynamics of the glassy background, which may result from either a single PNR or randomly-distributed coupled assembly of PNRs, we consider a system with $N_g$ degrees of freedom $\{x_\mu\}$ (Supplementary Note 13). 

	These position-like variables, related to the electric dipoles inside the PNRs, can be thought of as a multi-coordinate generalization of the usual single configuration coordinate ~\cite{Lang:1977p635,pelant2012luminescence,Kumar:2015p205117} for a defect or impurity in semiconductors. Such single coordinate defect model, though can give rise to very slow relaxation of photoresistivity~\cite{Lang:1977p635,pelant2012luminescence,Kumar:2015p205117}, is unlikely to lead to collective aging phenomena seen in our experiment.
 
	For our model, we thus assume a collective glassy background that gives rise to a two-step relaxation, $C_{gl}(t)=\langle x_\mu(t)x_\mu(0)\rangle=A \exp(-t/\tau_s)+B\exp[-(t/\tau_\alpha)^\beta]$ as a function of time $t$, for the dynamical correlation of $x_\mu$ at temperature $T$. The relaxation consists of a short-time exponential decay with the time scale $\tau_s$ and a long-time stretched exponential $\alpha$-relaxation with time scale $\tau_\alpha(T)$ and stretching exponent $\beta$ \cite{Reichman_2005, kob1997}. We assume that $\tau_s$ is temperature independent, whereas $\tau_\alpha$ increases with decreasing temperature. We also vary the coefficients $A(T)$ and $B(T)$ with $T$ such that the relative strength of the stretched exponential part increases at lower temperatures (Supplementary Note 13). In the above form of $C_{gl}(t)$, we neglect the $\beta$ relaxation ~\cite{kob1997} which leads to a power-law decay of $C_{gl}(t)$ in the plateau region ($\tau_s\ll t\ll \tau_\alpha$), after the microscopic relaxation and before the onset of the $\alpha$-relaxation. The $\beta$ relaxation is only expected to modify finer details of electronic relaxation in Fig.\ref{fig:5}.
	
	The glassy PNRs lead to a transition between the conduction band and impurity state via a coupling $H_{el-gl}=\sum_{i\alpha \mu}(V_{i\alpha \mu}c_i^\dagger f_\alpha +\mathrm{h.c})x_\mu$, where we take $V_{i\alpha \mu}$ as Gaussian complex random numbers to keep the model solvable. The coupling can arise either due to direct coupling of the electric field of the PNR to the electrons, or indirectly via coupling between PNR and electrons mediated by phonons, e.g., the soft polar optical phonon mode. 

	To gain an analytical understanding of the electron-hole recombination dynamics we consider a limit where there are no backactions of the conduction electrons on the impurity electrons and the glass (Supplementary Note 13).

	The description of the relaxation of the photo-excited electrons requires the consideration of the out-of-equilibrium quantum dynamics, which is beyond the scope of this paper. Instead, we consider an equilibrium dynamical correlation, namely the (connected) density-density correlation function $C_{el}(t)=\langle n_i(t)n_i(0)\rangle-\langle n_i(0)\rangle^2$ as a function of time $t$ for the density $n_i=c_i^\dagger c_i$ of the conduction electron. In our experiment, the resistivity decreases due to the increase of the density of conduction electrons via photo-excitations and relaxes through the relaxation of these excess carriers to the impurity states. The correlation function $C_{el}(t)$ characterizes a similar relaxation process, albeit close to the thermal equilibrium. The correlation function $C_{el}(t)$ can be computed exactly at a temperature $T$ in the toy model (Eq. \eqref{eq:Hamiltonian}). 

	In the limit of no backaction of the conduction electrons, 
	the electronic correlation function can be written as a convolution of the spectral function of the glass (Supplementary Note 13).
	As a result, the glass spectral function, which contains the information of the multiple time-scales and their non-trivial temperature dependence, may directly induce glassiness in the electronic relaxation.
	
	To verify the above scenario, we numerically compute $C_{el}(t)$ for two-step glass correlation functions $C_{gl}(t)$, shown in Fig. \ref{fig:5}(b) (upper panel) for three temperatures $T=0.5, 0.3, 0.1\ll \Delta$, where the temperature dependence is parameterized by $B(T)$ ($A=1-B$) and $\tau_\alpha(T)$ (Fig. \ref{fig:5}(b) (lower panel)). We take a temperature independent exponent $\beta=0.5$ and $\tau_\alpha(T)\sim T^{-2.8}$, consistent with our experimental results (Fig. \ref{fig:2}c). Here $\tau_s^{-1}$ ($\tau_s$) is set as the unit of energy (time) ($\hbar=1$). The calculated $C_{el}(t)$ is plotted in Fig. \ref{fig:5}(c) for a flat conduction band ($W=0$) (upper panel) and a semicircular conduction band DOS with $W=0.01$, and electron-glass coupling $V=0.3$. As shown in Fig. \ref{fig:5}(c), we find that for a local glassy bath, whose bandwidth is comparable or larger than the electronic energy scales $W$ and $\Delta$, the complex glassy correlation, namely the two-step relaxation with long and non-trivial temperature-dependent time scale, is also manifested in the electronic relaxation.

	\textbf{Outlook}:  While we do observe complex relaxations in such a toy model, these are still weak in contrast to the actual experimental results since our calculations only capture glassy two-step electronic relaxation via equilibrium dynamical correlations, whereas the actual experimental electron dynamics take place in a strongly non-equilibrium condition. We expect that a more realistic non-equilibrium theory considering the direct coupling with a real space distribution of PNRs embedded in the sea of conduction electrons would yield a strong glass. This would further demand intricate knowledge about the nature of interactions in the presence of PNRs which is still a subject of debate. Interestingly, these questions form the basis for understanding the nature of conduction in  an interesting class of materials known as polar metals~\cite{Shi:2013p1024,Wang:2019p61,Rischau:2017p643}, and hence we believe that our finding of glassy relaxations in presence of PNRs will be crucial in building the theory of conduction in quantum critical polar metals. Further, the observation of glassy dynamics deep inside the good metallic regime is in sharp contrast with the conventional semiconductors where glassy relaxation ceases to exist just before the insulator- metal transition (IMT) is approached from the insulating side. This raises the question about the envisaged role of glassy freezing of electrons as a precursor to IMT apart from the Anderson and Mott localization~\cite{Dobrosavljevi:2003p016402,dobrosavljevic:2012p}.

	\noindent\section*{Methods}
	\noindent\textbf{Sample preparation}: Oxygen deficient KTaO$_3$ single crystals were prepared by heating as received 001 oriented pristine  KTaO$_3$ substrate (from Princeton Scientific Corp.) in a vacuum sealed quartz tube in presence of titanium wire. For more details we refer to our previous work ~\cite{Ojha:2020p2000021,Ojha:2021p085120}.
	
	\noindent\textbf{Dielectric measurement}: Temperature-dependent dielectric measurement was performed in a close cycle cryostat using an impedance analyzer from Keysight Technology Instruments (Model No. E49908).
	
	\noindent\textbf{Transport measurement}: All the transport measurements were carried out in an ARS close cycle cryostat in van der Pauw geometry using a dc delta mode with a Keithley 6221 current source and a Keithley 2182A nanovoltmeter and also using standard low-frequency lock-in technique. Ohmic contacts were realized by ultrasonically bonding aluminum wire or by attaching gold wire with silver paint.
	
	\noindent\textbf{Light set-up}:
	Light of the desired wavelength was passed through the optical window of the close cycle cryostat from ARS. A home-built setup consisting of a diffuser and lens was used to make light fall homogeneously over the sample (see inset of Fig. \ref{fig:2}a ). Commercially available light-emitting diodes from Thor Labs were used as a light source. Incident power on the sample was measured with a laser check handheld power meter from coherent (Model No: 54-018).  
	
	\noindent\textbf{HAXPES measurement}: Near Fermi level and K 2p core level spectra were collected at Hard X-ray Photoelectron Spectroscopy (HAXPES) beamline (P22) of PETRA III, DESY, Hamburg, Germany using a high-resolution Phoibos electron analyzer~\cite{Schlueter:2019p040010}. Au Fermi level and Au 4f core level spectra collected on a gold foil (mounted on the same sample holder) were used as a reference for making the correction to the  measured kinetic energy. The chamber pressure during the measurement was $\sim$ 10$^{-10}$ Torr. An open cycle Helium flow cryostat was used to control the sample temperature. 
	
	\noindent\textbf{Second harmonic generation measurement}: SHG measurements were performed under reflection off the sample at a 45-degree incidence angle. A  $p$-polarised 800 nm beam from a Spectra-Physics Spirit-NOPA laser was used as the fundamental beam (pulse width: 300 fs, repetition rate: 1 MHz), and was focused on the sample surface. The $p$-polarised SHG intensity generated by the sample, was measured by a photo multiplier tube. The sample temperature was controlled by a helium-cooled Janis 300 cryostat installed with a heating element.

	\noindent\textbf{Density functional theory}: The noncolinear density functional theory calculations were carried out using the QUANTUM ESPRESSO package\cite{QE-2017}. In this calculation optimized norm-conserving pseudopotentials \cite{ONCV:PRB_Hamann085117, SCHLIPF201536, SOC:JCC1282016} were used and for the exchange-correlation functional \cite{PBE-GGA} we have incorporated Perdew, Burke and Ernzerhof generalized gradient approximation (PBE-GGA). For the unit cell, the Brillouin  zone was sampled with  $8 \cross 8 \cross 8$ $k$-points. The wave functions were expanded in plane waves with an energy up to 90 Ry. Since the effect of SOC in KTaO$_3$ is quite remarkable, we have employed full-relativistic pseudopotential for the Ta atom. The structural relaxations were performed until the force on each atom was reduced to 0.07 eV/\si{\angstrom}. The Berry phase calculations are carried out as implemented in the Wannier90 code~\cite{wannier90JOP:2020, MarzariRMP1419:2012}  
	We have used a $41 \times 41$ 2D k-mesh for the Berry curvature calculations. No significant change in the result is observed on increasing the k-mesh up to $101 \times 101$ (see Supplementary Note 8 for further details).

	\section*{Data availability}
	The authors declare that the data supporting the findings of this study are available within the main text and its Supplementary Information. Other relevant data are available from the corresponding author upon request.

	\section*{Acknowledgements}
	The work is funded by SERB, India, by a core research grant CRG/2022/001906 to SM. Portions of this research were carried out at the light source PETRA III DESY, a member of the Helmholtz Association (HGF). We would like to thank Dr. Anuradha Bhogra and Dr. Thiago Peixoto for their assistance at beamline P22. Financial support by the Department of Science \& Technology (Government of India) provided within the framework of the India@DESY collaboration is gratefully acknowledged. S.H. and V.G. acknowledge support from the US Department of Energy under grant no. DE-SC-0012375 for temperature-dependent second-harmonic generation measurements. SB acknowledges support from SERB (CRG/2022/001062), DST, India. SKG and MJ gratefully acknowledge Supercomputer Education and Research Centre, IISc for providing computational facilities SAHASRAT and PARAM-PRAVEGA. SKG acknowledges DST-Inspire fellowship (IF170557). SKO acknowledges the wire bonding facility at the Department of Physics, IISc Bangalore and thanks Shivam Nigam for the experimental assistance. SKO and SM thank Professor D. D. Sarma for giving them access to the quartz tube sealing and dielectric measurement setup in his lab.

	\section*{Author contribution}
	
	SM conceived and supervised the project. SKO, SM, and SK came up with all experimental plans to confirm the glassy behavior.  SKO, SH, JM carried out transport measurements. PM helped in making the photoconductivity setup. SH performed SHG measurements under the supervision of VG. SKO, PM, AG and CS carried out HAXPES measurements. SKO performed all the analysis of transport measurements and HAXPES.  SKG and MJ performed DFT calculations. S Bera, SK and SB provided phenomenological model calculations. 
	SKO, S Bera, SB, and SM wrote the manuscript with inputs from other authors.  All authors discussed the results and participated in finalizing the manuscript.

	\section*{Competing interests}
	The authors declare no competing interests.

	\begin{figure*}[htp]
		\centering{
			{~}\hspace*{-0.2cm}
			\includegraphics[scale=.85 ]{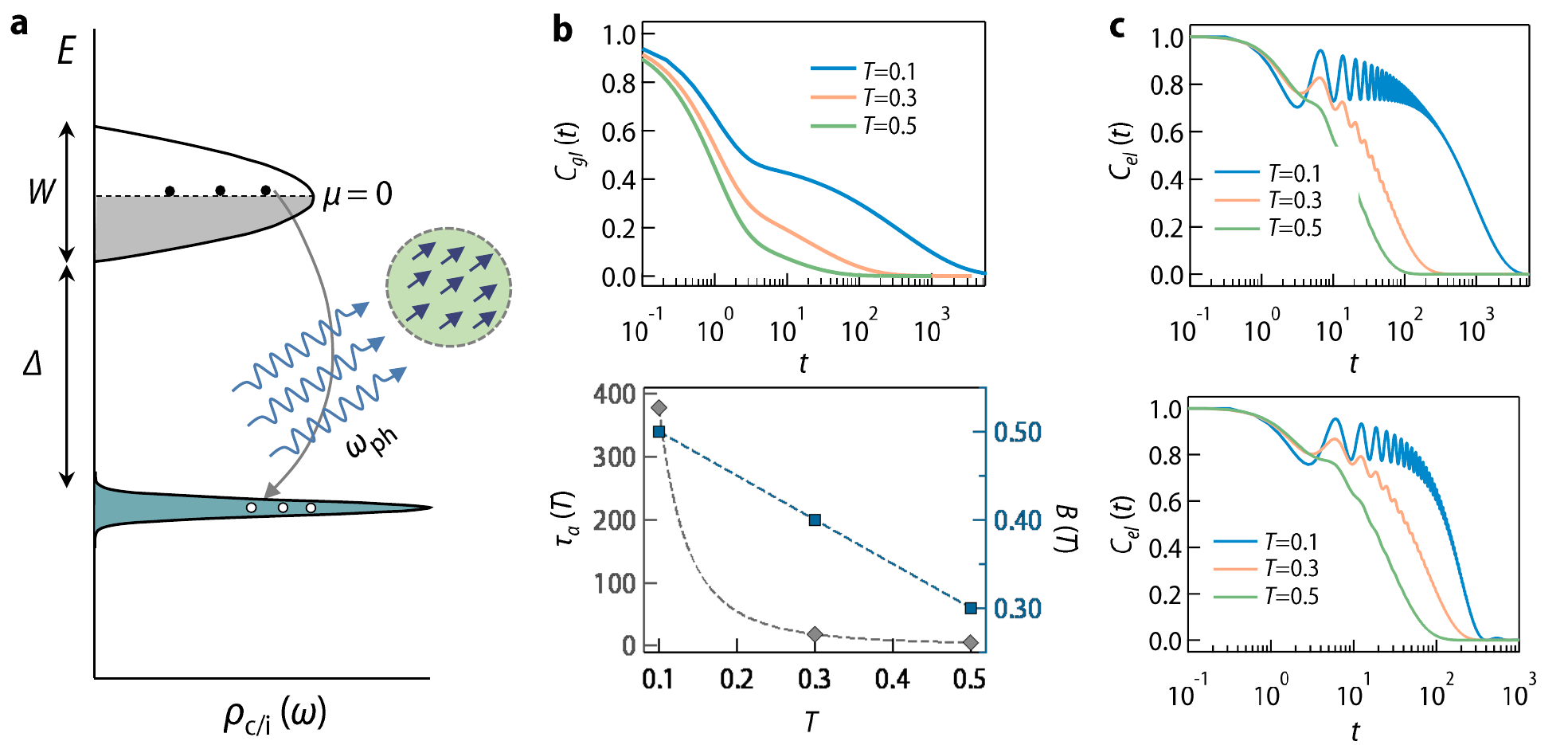}
			\caption{\textbf{Theoretical calculations} \textbf{a.} A schematic of band diagram considered for computing inter band transitions when the conduction electron is coupled to glassy background from randomly oriented polar nano regions. \textbf{b.}  (Upper panel) The glass correlation function $C_{gl}(t)$ vs. $t$ for three temperatures for $\tau_s(T)=1$, $b=0.5$. (Lower panel) Plot of $\tau_\alpha (T)\sim T^{-2.8}$ vs. $T$ (left y axis, in units of $\tau_s$)  and $B(T)$ (right y axis)  vs. $T$ ($A=1-B$) for three temperatures. \textbf{c.}  (upper panel) The density-density correlation function of conduction electron $C_{el}(t)$ vs. $t$ for three temperatures for flat ($W=0$) conduction band. (lower panel) $C_{el}(t)$ vs. $t$ for three temperatures for semicircular conduction band with $W=0.01$. Source data are provided as a Source Data file.} \label{fig:5}}
	\end{figure*}

\clearpage
\newpage
\renewcommand{\figurename}{\textbf{Supplementary Figure}}
\setcounter{figure}{0}

\hspace{5cm}\textbf{\Large{Supplementary Information}}

\vspace{1cm}
\noindent\textbf{List of contents}
\vspace{0.3cm}

\noindent\textbf{Note 1.} Calculation of electron's mean free path.

\noindent\textbf{Note 2.} Sheet resistance vs temperature plot.

\noindent\textbf{Note 3.} Stretched exponential behavior of resistance relaxation in the light off stage.

\noindent\textbf{Note 4.} Power law behavior of persistence photo-resistance relaxation time ($\tau$).

\noindent\textbf{Note 5.} Experiment with red light.

\noindent\textbf{Note 6.} Long-time relaxation of persistence photo-resistance.

\noindent\textbf{Note 7.} Deviation of persistence photo-resistance relaxation time ($\tau$) from activated behavior at low temperature.

\noindent\textbf{Note 8.} Dielectric loss in pristine KTaO$_3$.

\noindent\textbf{Note 9.} Correspondence between the appearance of polar nano regions and photo-doping effect.

\noindent\textbf{Note 10.} Raman measurement on a metallic oxygen deficient KTO.

\noindent\textbf{Note 11.} Presence of potassium vacancy.

\noindent\textbf{Note 12.} Berry phase calculations.

\noindent\textbf{Note 13.} Toy model of complex inter-band electronic relaxation mediated by a glassy bath.

\clearpage

\clearpage

\noindent\textbf{Supplementary Note 1: Calculation of electron's mean free path}

\vspace{0.5cm}

The Fermi surface of the electron doped KTO~\cite{Mattheiss:1972p4718} comprises of three ellipsoids of revolution centered at $\Gamma$ point (see Supplementary Fig. 1a) with major axis $k_{F,max}$ and the minor axis ${k_{F,min}}$ transverse to it. In the presence of magnetic field along [001] crystallographic axis, electrons traverse around the extremal orbits in the momentum space as shown in the Supplementary Fig. 1b, leading to the observation of SdH oscillations in magnetoresistance. Please refer to our earlier work~\cite{Ojha:2020p2000021} for SdH oscillation data on the same sample which is being investigated in the current study. These oscillations are periodic in 1/$B$ whose frequency $F$$_\textnormal{SdH}$ is related to the area of extremal orbits A$_{ext}$ as given below~\cite{Herranz:2007p216803}

\begin{figure}[ht]
	\vspace{-0pt}
	\centering{
		\includegraphics[width=0.8\textwidth] {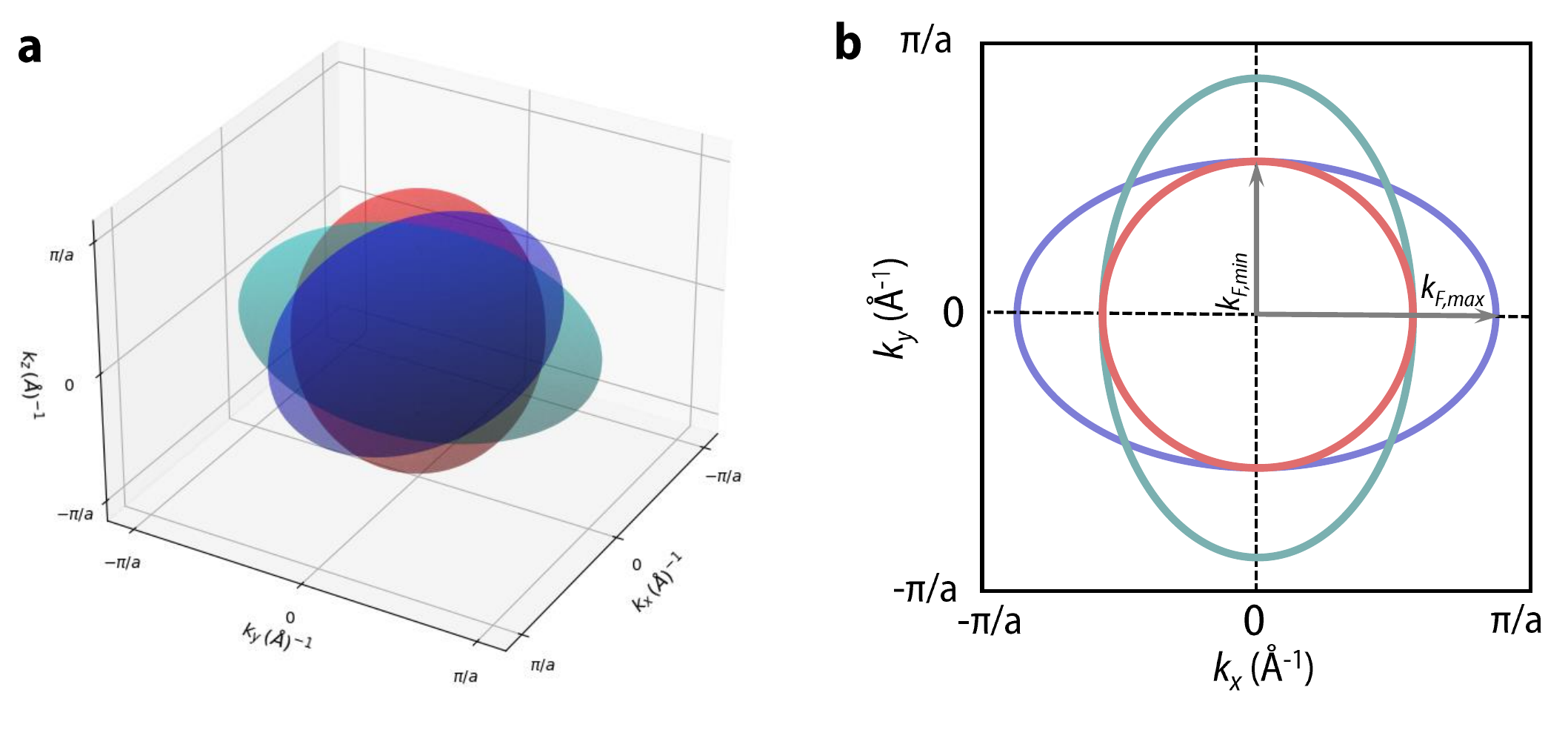}
		\caption{\textbf{a.} Ellipsoidal Fermi surfaces of electron doped KTO. Here $k_{F,max}$ was taken to be 1.541 times $k_{F,min}$ for plotting. \textbf{b.} Cross section of the Fermi surface with the plane $k_Z$=0}\label{GFig9}}
\end{figure} 

\begin{equation}
	F_\textnormal{SdH}=\hbar A_{ext}/{2\pi e} \label{eq1}
\end{equation}

The extremal orbits corresponding to the two ellipsoids directed along $k_x$ and $k_y$ axis are ellipses with  area equal to $\pi$$k_{F,min}$${k_{F,max}}$, whereas the extremal orbit corresponding to the ellipsoid along the $k_z$ direction is circular with an area ${\pi k^{2}_{F,min}}$ (see Supplementary Fig. 1b). Further, from the literature it is known that~\cite{Uwe:1979p3041,Mattheiss:1972p4718}

\begin{equation}
	k_{F,max}/k_{F,min}=1.541 \label{eq2}
\end{equation}

for electron doped KTO, which would further imply that there are more electrons around elliptical orbit than those orbiting around circular orbit~\cite{Herranz:2006p064403}. From above fact, we assume that the main frequency of SdH oscillation comes from electrons orbiting around the elliptical orbit and therefore $A_{ext}$ in Supplementary Eq. (1) is taken to be the area of the ellipse $\pi$$k_{F,min}$${k_{F,max}}$. From the SdH analysis,  $F$$_\textnormal{SdH}$ for our oxygen deficient KTO is found to be 12.8 Tesla~\cite{Ojha:2020p2000021}. Putting this value in Supplementary Eq. (1) and using the Supplementary Eq. (2), we first calculate $k_{F,max}$ and $k_{F,min}$ individually and then 3D  carrier density is determined. For this we note that the total volume occupied by three ellipsoid is V${_k=3\times(4\pi/3){k^2}_{F,min}k_{F,max}}$ from which the carrier concentration $n$ can be calculated as~\cite{Herranz:2006p064403}

\begin{equation}
	n={({k^2}_{F,min}k_{F,max})/{\pi^2}} \label{eq3}
\end{equation}

Putting the value of $k_{F,max}$ and $k_{F,min}$ in the above Supplementary Eq. (3), the 3D carrier density $n$ for the oxygen deficient KTO is found to be 5.7$\times$10$^{17}$ cm$^{-3}$. Having obtained the 3D carrier density, we next compute the temperature dependent electron's mean free path following the well established approach described in the papers~\cite{Lin:2017p41,Collignon:2020p031025}. 

In the Drude-Boltzmann picture, 3 dimensional conductivity ($\sigma$) of a metal with spherical Fermi surface is given by

\begin{equation}
	\sigma=ne^2\tau/m^*=e^2(k_F)^2l_e/3\pi^2\hbar \label{eq4}
\end{equation}

\noindent  where $n$ is the 3D carrier density given by $n$=$k_F$$^3$/3$\pi$$^2$  ($k_F$ is the Fermi wavevector), and $\tau$ is scattering time constant given by $\tau$=$l_e$$m^*$/$\hbar$$k_F$ where $m$$^*$ is the effective mass of electrons and $l_e$ is the electron's mean free path~\cite{ashcroft2022solid}. Rearranging the Supplementary Eq. (4), the expression for $l_e$ can be written as

\begin{equation}
	l_e=3\pi^2\hbar/\rho(k_F)^2 e^2 \label{eq5}
\end{equation}

\noindent  where $\rho$=1/$\sigma$ is the resistivity. Since all the measurements in the present study are performed in Van der Pauw geometry, the temperature dependent $\rho$ is obtained from the formula

\begin{equation}
	\rho=(\pi/ln2)Rt \label{eq6}
\end{equation}

\noindent where $R$ is the measured temperature dependent resistance and $t$ is the thickness of the conducting region. In the present case, $t$ is calculated by equating the 3D carrier density obtained from the SdH analysis with the sheet carrier density $n_S$ obtained from the Hall measurement divided by the thickness [$n$=$n_S$/$t$, see the reference ~\cite{Herranz:2007p216803}]. Apart from the $\rho$,  $k_F$ is the another input parameter in Supplementary Eq. (5) for calculating $l_e$. Since the Supplementary Eq. (5) holds only for the spherical Fermi surface, we make the following approximation. Since the Fermi surface of KTO is only moderately anisotropic at such dilute carrier density, for simplification, the value of $k_F$ has been calculated by effectively mapping the ellipsoidal Fermi surface onto a spherical Fermi surface wherein we equate the total volume of the three ellipsoids with a single spherical Fermi surface of radius $k_F$. The value of $k_F$ is then estimated from the relation $k_F$=(3$\pi$$^2$$n$)$^{1/3}$~\cite{ashcroft2022solid}. Once we obtain $k_F$, the temperature dependent $\rho$ obtained from the Supplementary Eq. (6) along with $k_F$ is plugged in the Supplementary Eq. (5) and temperature dependent electron's mean free path is calculated.

$l_e$ calculated using this approach for the oxygen deficient KTO has been shown by a blue curve in the Fig. 1d of the main text. As evident from the plot, above $\sim$130 K, $l_e$ becomes shorter than the inverse of $k_F$ ($k_F$$l_e$ $>$ 1) and the system enters into a bad metal phase. A further correction to the calculated $l_e$ is required above the Fermion degeneracy temperature above which the thermal de Broglie wavelength (given by $\Lambda$=$h$/$\sqrt{2\pi m^*k_BT}$) becomes larger than the inter electron separation given by $n$$^{-1/3}$. Above this temperature, electrons become non-degenerate and the modified $l_e$ is given by $l'_e$$\sim$ $l_e$ $n$$^{-1/3}$/$\Lambda$~\cite{Lin:2017p41,Collignon:2020p031025}. For calculation of $l'_e$, one requires an estimation of $m^*$. In the present work we estimate the  $m^*$ by analyzing the  temperature  dependent SdH oscillations and is found to be 0.56 $m_e$ (please refer to the section B of the supplemental material of our earlier paper~\cite{Ojha:2020p2000021} for more details). The orange curve in Fig. 1d of the main text shows the calculated $l_e$ for the non-degenerate case. As evident, this exercise only shifts the curve to a little higher temperature and crossover to bad metal phase occurs around 145 K, however, the conclusions remain the same. In the current work, observation of glassy dynamics is inherently constrained to the temperatures which is much lower than the crossover temperature to bad metal phase and hence for all practical purposes our electron doped KTO system is in good metal regime with well defined scattering.

\clearpage

\noindent\textbf{Supplementary Note 2: Sheet resistance vs temperature plot.}

\vspace{0.5cm}

As depicted in Fig. 2b of the main text, shining light above 150 K does not have any noticeable effect. This observation is further supported by the plot shown in Supplementary Fig. 2, where we compare the $R_S$ vs $T$ curve obtained in the dark (taken in the cooling run) with the measurement conducted during the heating run after exposing the sample to green light (power= 145 $\mu$ Watt) for 30 minutes at 10 K. It is evident from the plot that the two curves converge around 150 K, indicating a lack of any significant photo-doping above 150 K.

\begin{figure}[h]
	\centering{
		\includegraphics[scale=0.85]{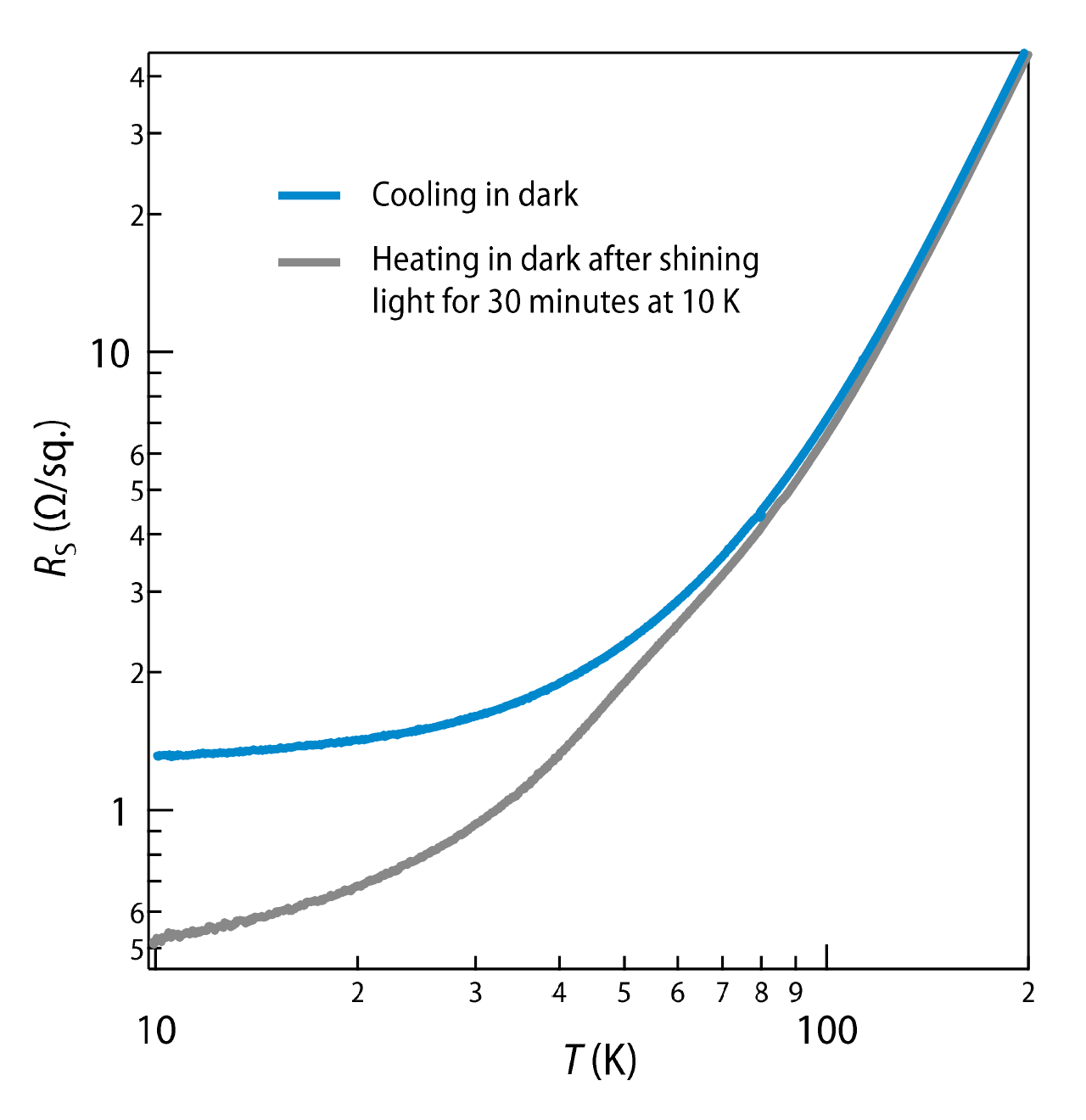}
		\caption{Temperature dependent sheet resistance ($R$$_{\text{S}}$) of oxygen-deficient KTaO$_3$ sample ($n$=5.7$\cross$10$^{17}$ cm$^{-3}$) taken in a cooling run in dark (sky blue curve). The slate-gray color curve shows  data taken in a heating run after shining light for 30 minutes at 10 K.}
		\label{Fig2}}
\end{figure}

\clearpage

\noindent\textbf{Supplementary Note 3: Stretched exponential behavior of resistance relaxation in the light off stage.}

\vspace{0.5cm}
The panels a-f in the Supplementary Fig. 3 show the temporal evolution of resistance before and after turning off the light along with the fitting in light off stage with an stretched exponential function (exp(-($t$/$\tau$)$^\beta$) where $\tau$ is the relaxation time and $\beta$ (stretching exponent) $<$ 1). As evident, this function provides an excellent fit to experimental data at range of temperatures. 

\begin{figure}[h]
	\centering{
		\includegraphics[scale=0.55]{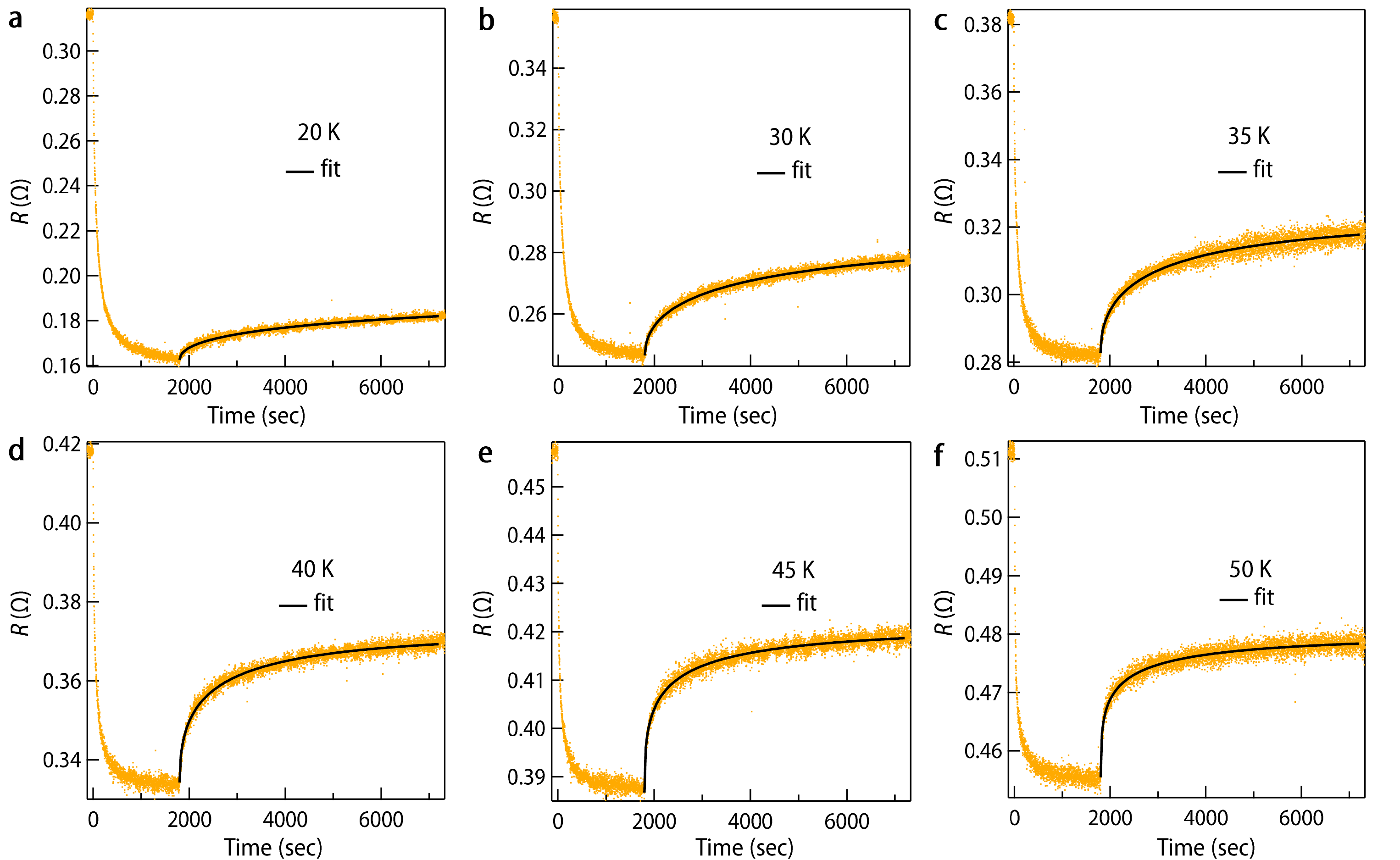}
		\caption{Fitting of resistance relaxation in the light off stage with stretched exponential function at several fixed temperatures.}
		\label{Fig_fit}}
\end{figure}

\clearpage

\noindent\textbf{Supplementary Note 4: Power law behavior of persistence photo-resistance relaxation time ($\tau$).}
\vspace{0.5cm}

Supplementary Fig. 4 shows $log$-$log$ plot of relaxation time ($\tau$) as a function of temperature. As evident, the plot looks linear in wide range of temperatures belo 50 K signifying a power law dependence ($\tau\sim T^{-\alpha}$). The value of alpha obtained from fitting is found to be 2.8. We also emphasize that such power-law divergence of relaxation time  has been also discussed theoretically in context of $\alpha$ relaxation in glasses~\cite{kob2002supercooled}.

\begin{figure}[h]
	\centering{
		\includegraphics[scale=0.55]{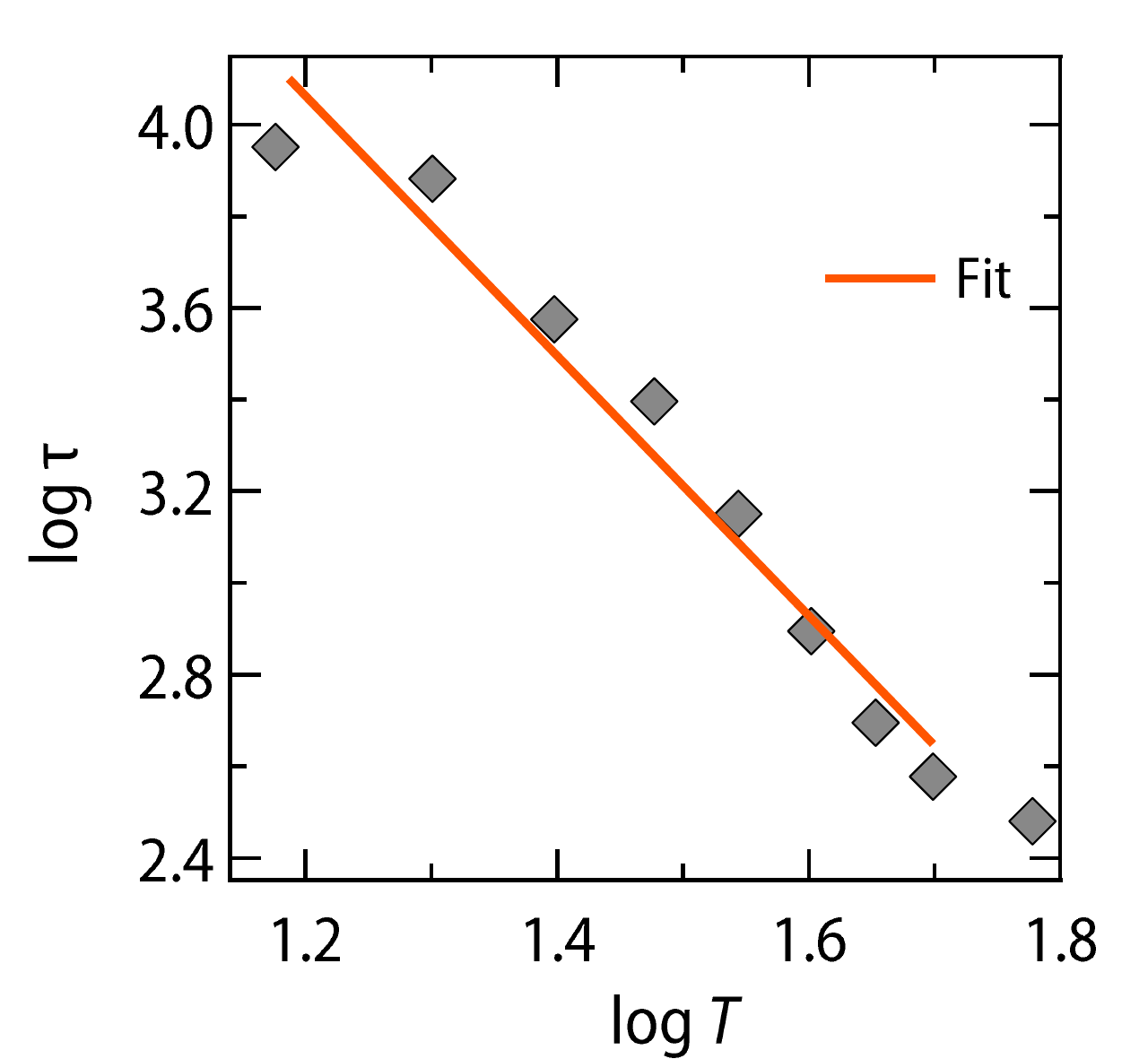}
		\caption{Power law temperature dependence of relaxation time ($\tau$).}
		\label{Fig_pow}}
\end{figure}

\clearpage

\noindent\textbf{Supplementary Note 5: Experiment with red light.}

\vspace{0.5cm}

In the main text, we have presented all the measurements conducted using the green light. In this section, we provide an additional set of data obtained in three consecutive cycles (see Supplementary Fig. 5a) using a red light of wavelength $\lambda$ = 650 nm, power = 60 $\mu$ Watt). Supplementary Fig. 5b and 5c shows the temperature evolution of $\beta$ and $\tau$ respectively obtained from the fitting. As evident, while the enhancement in $\tau$ below 50 K is consistent with results with green light, $\beta$ slightly decreases from its peak value at lower temperatures. However, this is within the error bar and the results qualitatively align with the findings from the measurements using green light and suggest that the behavior observed is not specific to a particular wavelength of light.

\begin{figure}[h]
	\centering{
		\includegraphics[scale=0.43]{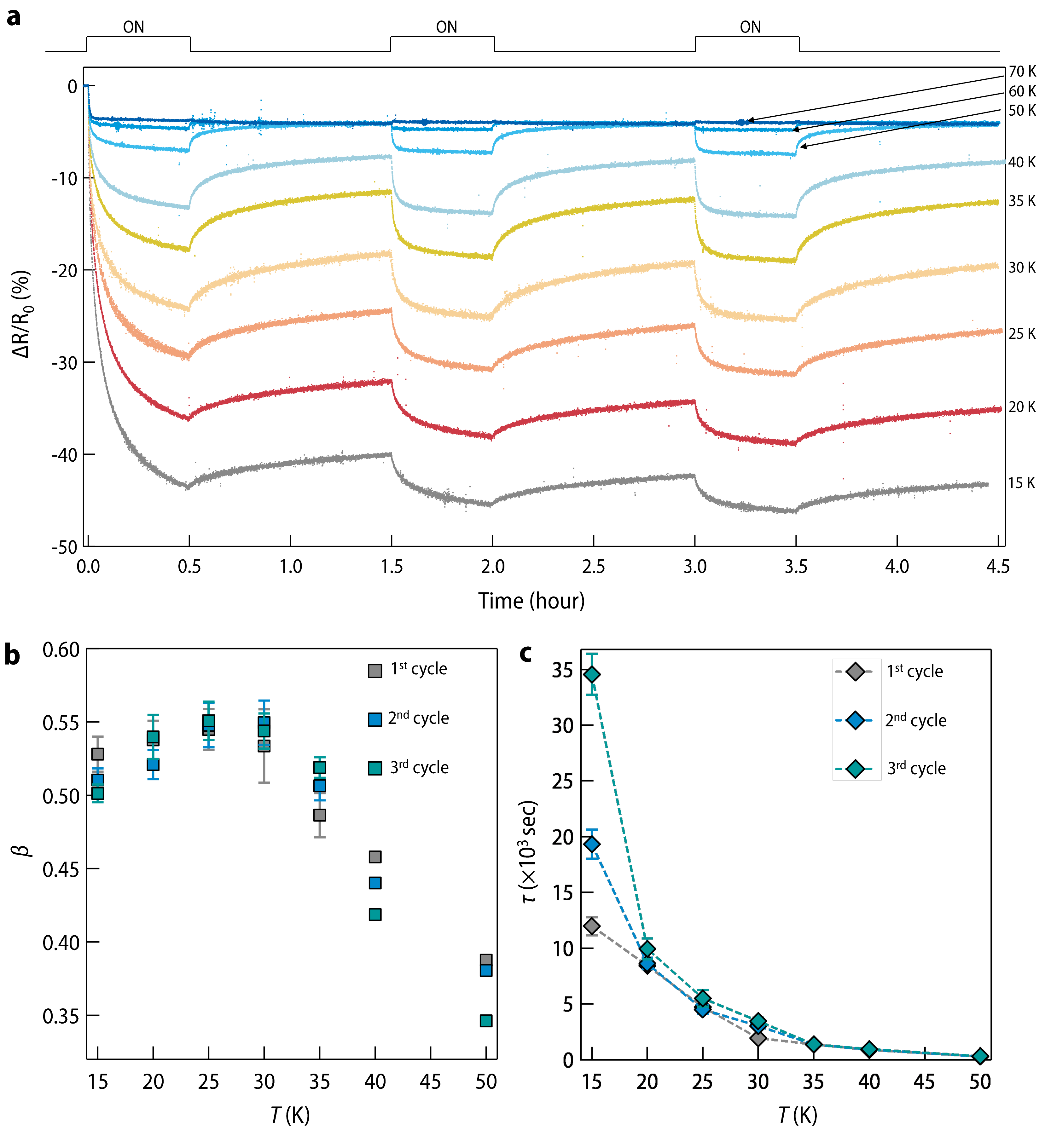}
		\caption{\textbf{a.} Temporal evolution of resistance under red light illumination ($\lambda$ = 650 nm, power = 60 $\mu$ watt) for 30 minutes measured at several fixed temperatures. After 30 minutes, resistance relaxation was observed in dark for the next one hour. This measurement was repeated for 3 consecutive cycles. For comparative analysis, change in resistance has been converted into relative percentage change ($\Delta$$R$/$R_0$)$\cross$100. \textbf{b.} Temperature dependence of  stretching exponent ($\beta$) obtained from fitting of resistance relaxation in light off stage with a stretched exponential function for all three cycles. \textbf{c.} Temperature dependence of  corresponding relaxation time ($\tau$) for all three cycles.}
		\label{Fig3}}
\end{figure}

\clearpage

\noindent\textbf{Supplementary Note 6: Long-time relaxation of persistence photo-resistance.}

\vspace{0.5 cm}

In Fig. 2c of the main text and Supplementary Fig. 5b, the value of the stretching exponent  was determined by fitting the resistance relaxation during the off-stage, which lasted for 1.5 hours. To ensure that the resistance relaxation was substantial enough to yield a reliable fitting, we conducted an additional measurement where the resistance relaxation was observed for an entire day (see Supplementary Fig. 6). It is important to emphasize that even when the data is fitted for an extended period of up to 24 hours, the same value of $\beta$ = 0.5 is obtained.

\begin{figure}[h]
	\centering{
		\includegraphics[scale=0.5]{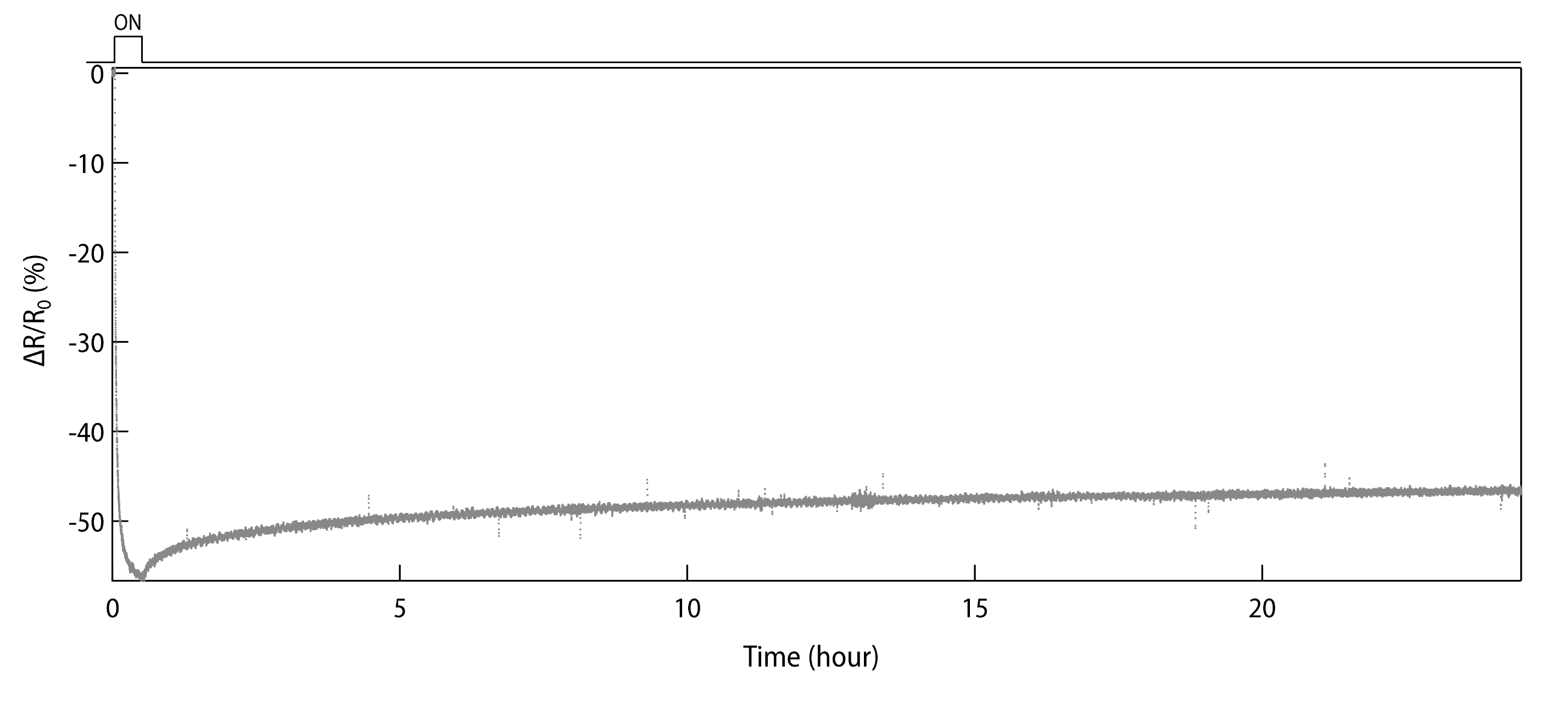}
		\caption{Temporal evolution of resistance under green light illumination ($\lambda$ = 527 nm, power = 145 $\mu$ Watt) for 30 minutes measured at 15 K. After 30 minutes, resistance relaxation was observed in dark for the next 24 hours. For comparative analysis, change in resistance has been converted into relative percentage change ($\Delta$$R$/$R_0$)$\cross$100.  }
		\label{Fig4}}
\end{figure}

\clearpage

\noindent\textbf{Supplementary Note 7: Deviation of persistence photo-resistance relaxation time ($\tau$) from activated behavior at low temperature.}

\vspace{0.5 cm}
In the large lattice relaxation (LLR) model~\cite{Lang:1977p635}, the recombination of electron-hole pairs occurs through the thermal excitation of electrons over an energy barrier. This process is  an activated process of the Arrhenius type. To validate this model, we have plotted ln $\tau$ against 1000/$T$ in Supplementary Fig. 7. It is evident from the plot that the data does not exhibit the expected linear behavior at low temperatures signifying that the LLR model can not account for our experimental observation of glassy dynamics below 35 K.

\begin{figure}[h]
	\centering{
		\includegraphics[scale=0.8]{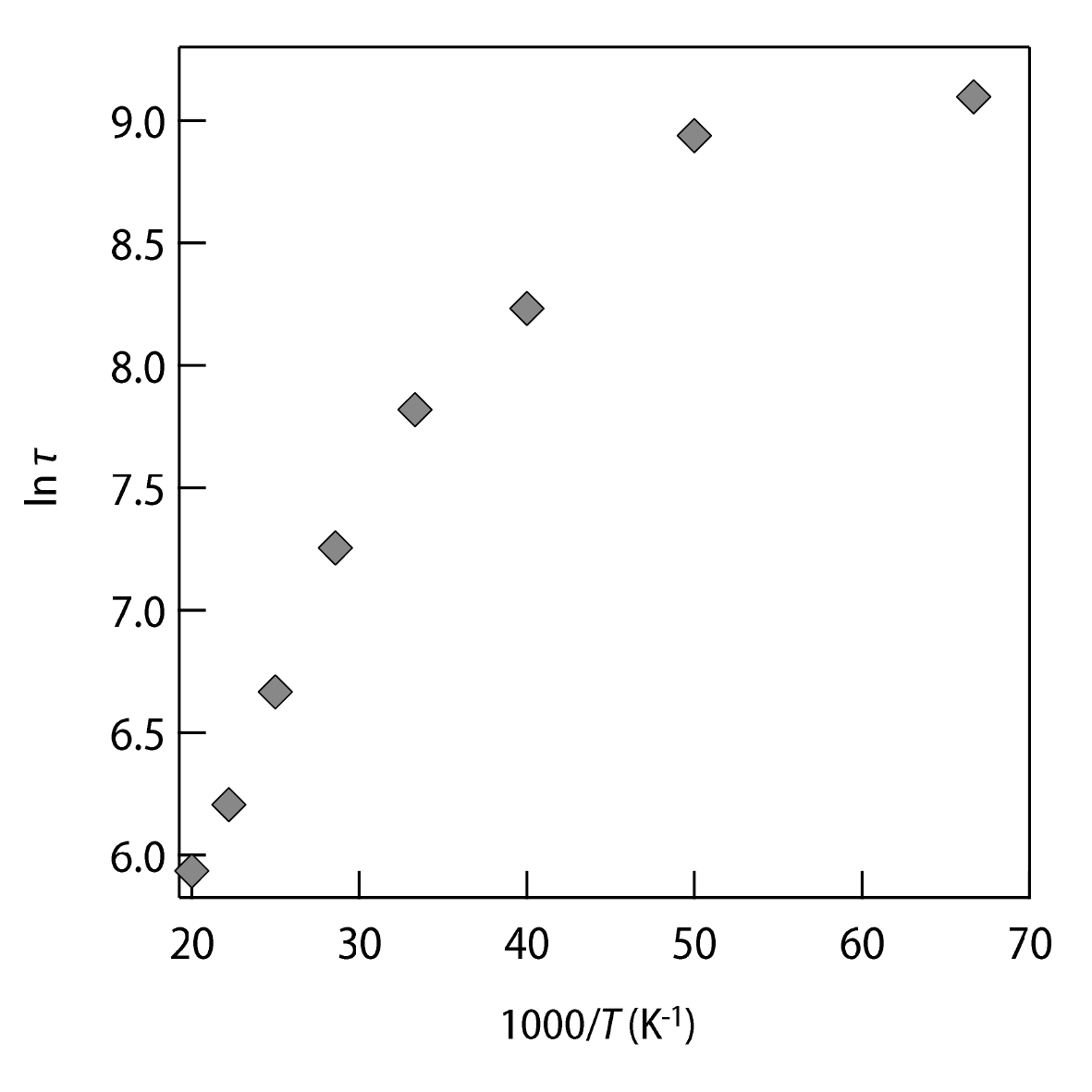}
		\caption{ Arrhenius plot of photo-resistance relaxation time ($\tau$). As evident, a clear deviation from activation behavior is observed below 35 K. }
		\label{Fig5}}
\end{figure}

\clearpage

\noindent\textbf{Supplementary Note 8: Dielectric loss in pristine KTaO$_3$.}

\vspace{0.5cm}

In its ideal form, KTaO$_3$ is perfectly centrosymmetric and hence, should not possess any electric dipole moment. However, the presence of impurities and disorder can disrupt the crystal's inversion symmetry, leading to the development of permanent electric dipoles~\cite{Grenier:1989p2515,Geifman:1997p115,Voigt:1994p853,Grenier:1992p105,Trybula:2015p23,Tkach:2021p1222}. In highly polarizable host such as KTaO$_3$, these dipoles polarize the surrounding lattice leading to formation of polar nano regions (PNRs)~\cite{Samara:2003pR367}. Under an applied AC electric field, these PNRs act as a source of dielectric losses, which can be observed as a peak in the loss tangent (tan $\delta$ = $\epsilon$$^{''}$/$\epsilon$$^{'}$, where $\epsilon$$^{'}$ and $\epsilon$$^{''}$ are the real and imaginary parts of the complex dielectric function, respectively).

Our temperature-dependent measurement of dielectric function indeed reveals the presence of PNRs even in our pristine KTaO$_3$ single crystal (see inset of Supplementary Fig. 8). Further, frequency-dependent measurements reveal that the dielectric loss is an activated process emphasizing that the PNRs in pristine crystals are very dilute and independent.

\begin{figure}[h]
	\centering{
		\includegraphics[scale=0.8]{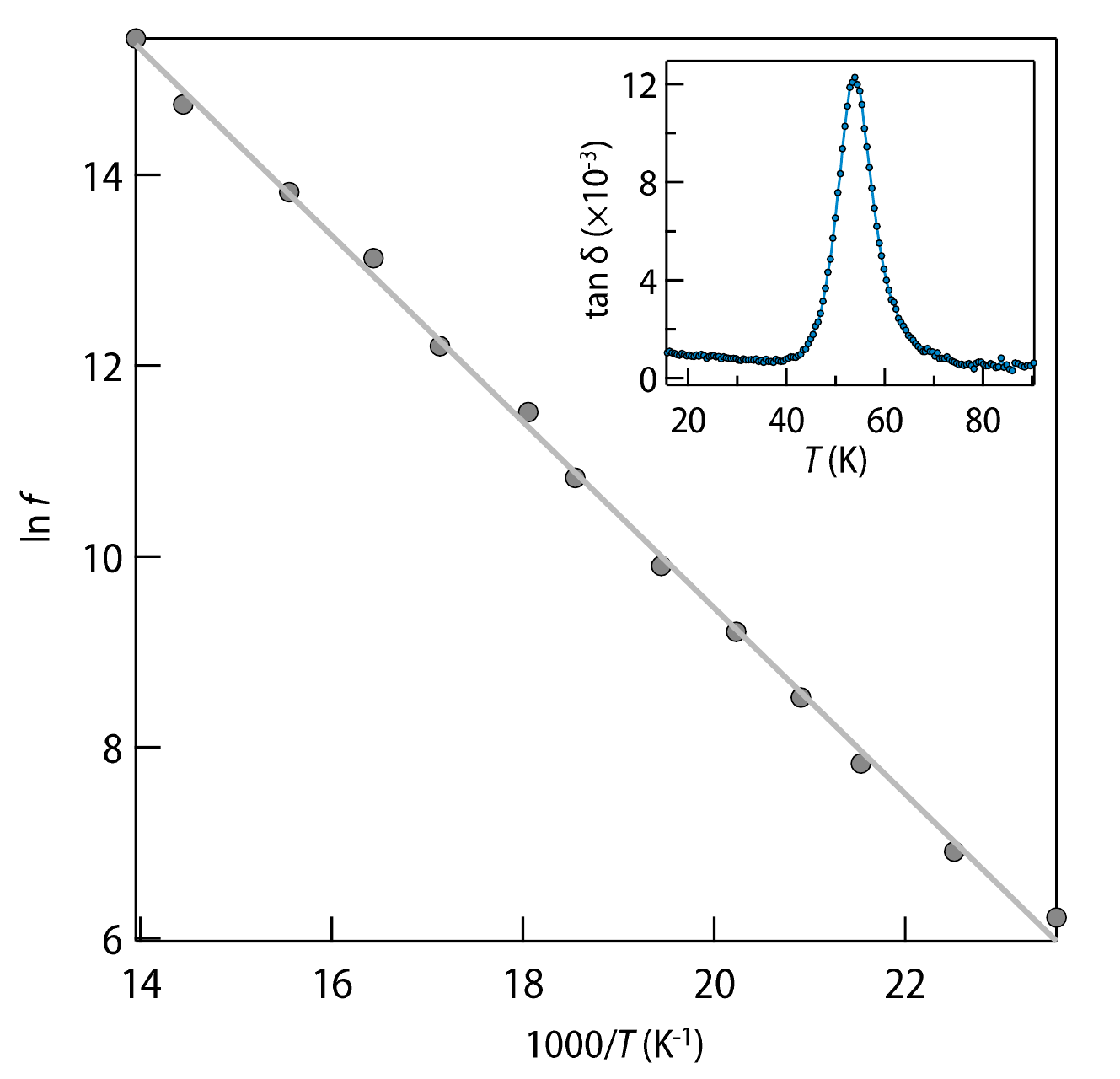}
		\caption{ Arrhenius plot of dielectric relaxation frequency ($f$) for pristine KTaO$_3$. Inset shows the dielectric loss tangent (tan $\delta$) around  54 K at $f$ = 50 kHz.}
		\label{Fig1}}
\end{figure}

\clearpage

\noindent\textbf{Supplementary Note 9. Correspondence between the appearance of polar nano regions and photo-doping effect.}

\vspace{0.5 cm}

In order to study the correlation between PNRs and photo-doping effect, we have compared the temperature-dependent total SHG intensity with the fraction of resistance (in terms of relative percentage change ($\Delta$$R$/$R_0$)$\cross$100) which has not been recovered at the end of 1.5 hours after turning off the light. We call this quantity the persistence photo resistance denoted with the symbol $\delta$. See the inset of the bottom panel of Supplementary Fig. 9 for the definition of $\delta$.  As evident, the appearance of finite signal in SHG exactly coincides with the $\delta$, strongly signifying the direct role of PNRs behind the effective electron-hole separation of our samples.

\begin{figure}[h]
	\centering{
		\includegraphics[scale=0.65]{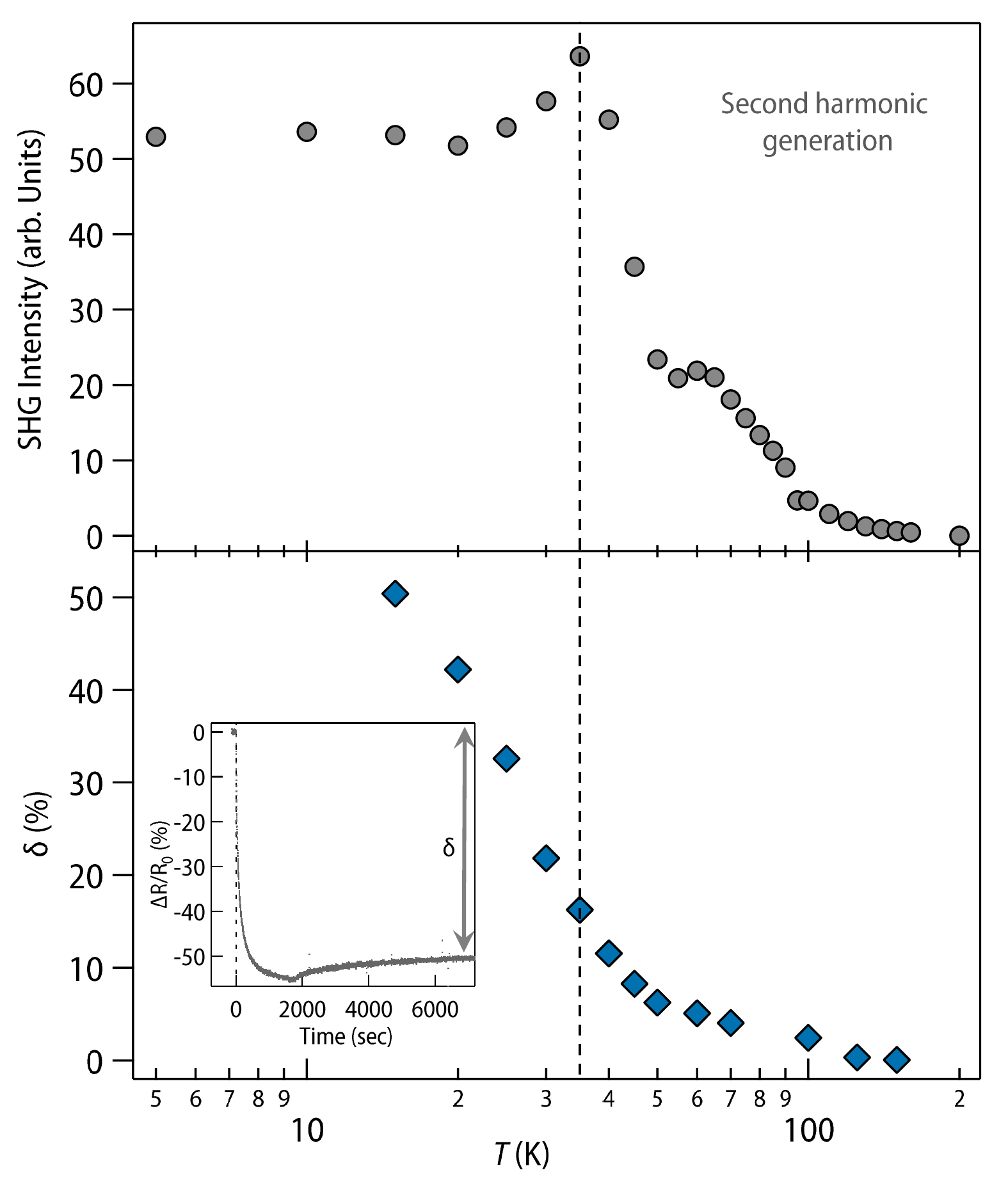}
		\caption{ (upper panel) Temperature-dependent SHG intensity measured on oxygen deficient KTaO$_3$ sample. (lower panel) Temperature dependence of persistence photo resistance ($\delta$) in terms of relative percentage change at the end of 1.5 hours after turning off the light (see the inset of the lower panel for the definition of $\delta$).}
		\label{Fig8}}
\end{figure}

\clearpage

\noindent\textbf{Supplementary Note 10: Raman measurement on a metallic oxygen deficient KTO.}

\vspace{0.5cm}

Supplementary Fig. 10 shows the Raman spectra for one of the metallic oxygen-deficient KTO sample at room temperature. As evident, the soft polar mode (TO$_1$, marked by a red arrow) is preserved even in the metallic sample~\cite{Shirane:1967p396,Nilsen:2004p1413,Perry:2004p1619,Uwe:1986p6436,Golovina:2012p614}. This result is consistent with our SHG measurement.

\begin{figure}[h]
	\centering{
		\includegraphics[scale=0.55]{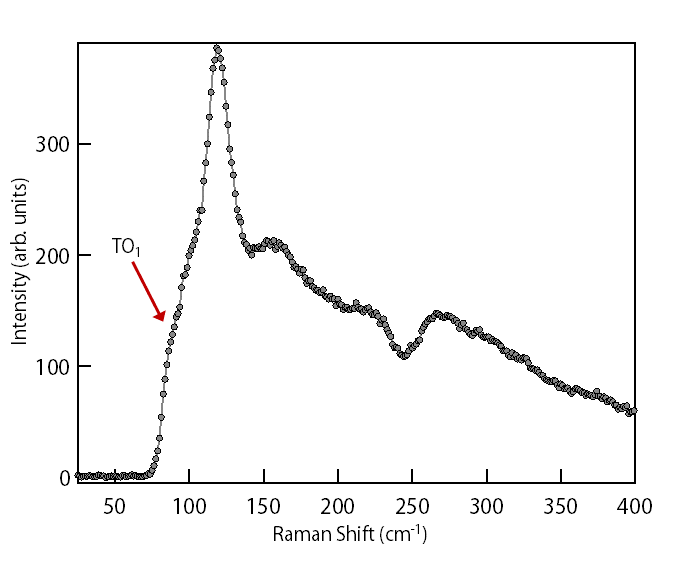}
		\caption{Raman spectra of a metallic oxygen deficient KTO sample at room temperature.}
		\label{Fig_Raman}}
\end{figure}

\clearpage

\noindent\textbf{Supplementary Note 11: Presence of potassium vacancy.}
\vspace{0.5cm}

To investigate the presence of potassium vacancy in our oxygen-deficient KTaO$_3$ sample, K 2p core levels were collected at the Hard X-ray Photoelectron Spectroscopy (HAXPES) beamline (P22) at PETRA III, DESY.  Supplementary Fig. 11a shows one representative experimental data (recorded with an incident photon energy of 5800 eV at room temperature) along with its fitting with the convolution of Lorentzian and Gaussian functions. It is evident from the figure that, in addition to the K 2p 3/2 peak at 292.65 eV arising from the lattice, there is an extra peak appearing at 293.1 eV. It has been previously shown that the presence of an additional peak at a higher binding energy is a characteristic feature of a potassium vacancy in the system~\cite{Kubacki:2012p1252}.
In order to further understand the potassium vacancy profile in our sample, we have carried out measurements with varying photon energy from 3400 eV to 5800 eV (see Table I for the values of mean free path (MFP) and mean escape depth (MED)). 
As evident from Supplementary Fig. 11b, there are hardly any significant alterations in the line shape of the spectra as the photon energy was increased. This observation indicates that the potassium vacancy profile is homogeneous throughout the bulk of the sample.

\begin{figure}[h]
	\centering{
		\includegraphics[scale=0.50]{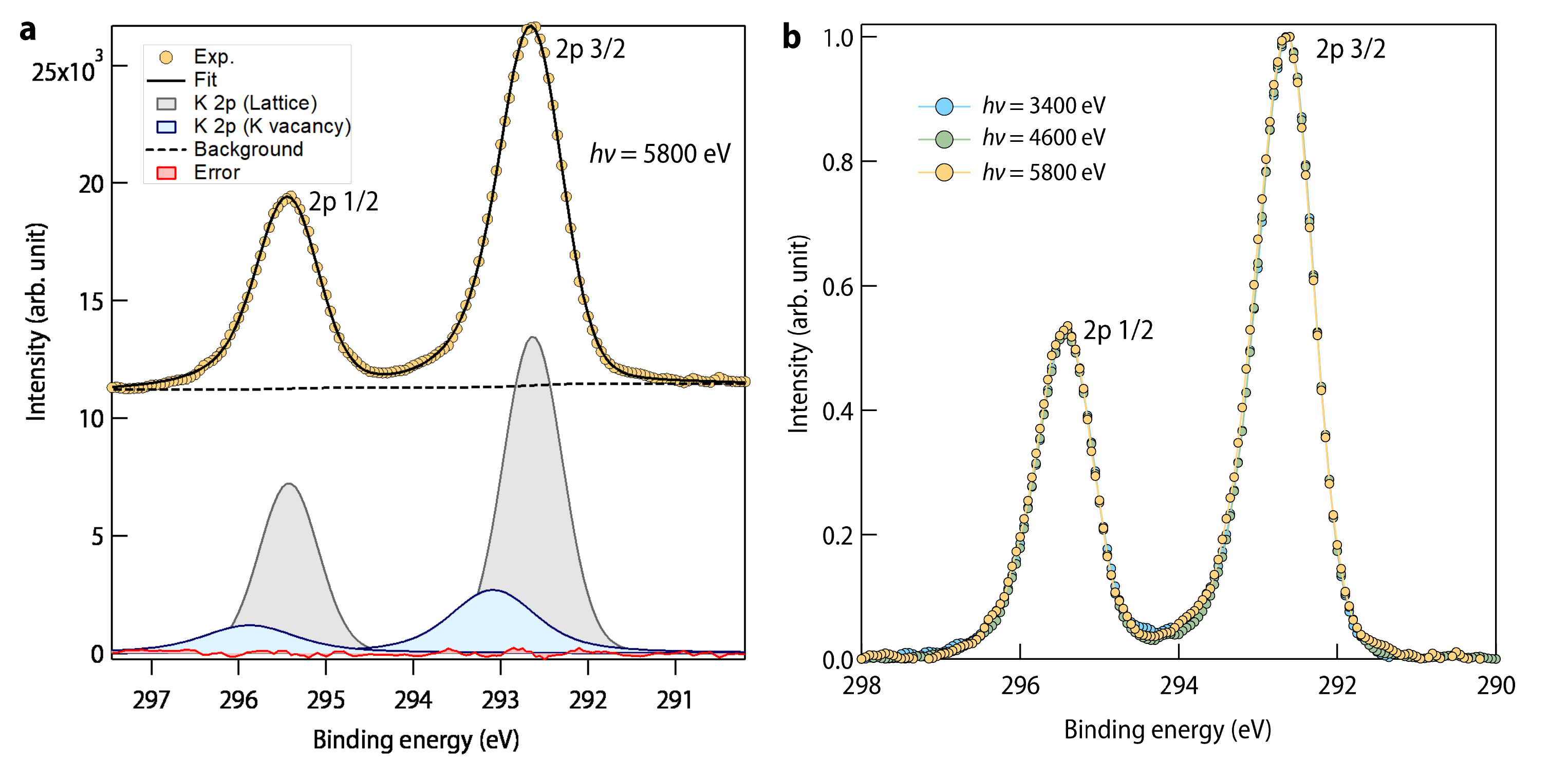}
		\caption{\textbf{a} Deconvolution of K 2p core level spectrum (for oxygen-deficient KTaO$_3$ sample) by using the convolution of Lorentzian and Gaussian function. This data was recorded at room temperature with photon energy $h$$\nu$ = 5800 eV. Yellow-filled circles denote the experimental data and a solid black line denotes the simulated spectra. Curves filled with gray and sky blue color correspond to the lattice potassium and potassium vacancy respectively. The dashed black line denotes the Shirley background and the red-filled curve corresponds to the difference between the experimental data and simulated spectra.  \textbf{b} K 2p core level spectra recorded with varying incident photon energy. All the spectra were recorded at room temperature.}
		\label{Fig7}}
\end{figure}

\begin{table}[h]
	\begin{center}
		\begin{tabular}{c|c|c} 
			\hline
			\text{Photon energy (eV)} & \text{MFP (nm)} & \text{MED (nm)} \\
			
			\hline
			3400  &   5  &  15\\
			4600  &   6.4  &  19.2\\
			5800  &   7.7  &  23.1\\
			\hline
		\end{tabular}      
		\caption{Table of mean free path (MFP) and mean escape depth (MED) with increasing photon energy. MFP was calculated from the formula $m$(K.E.)$^{\gamma}$ where K.E. is the kinetic energy of the ejected electrons and the values of $m$ and $\gamma$ were taken to be 0.12 and 0.75 respectively~\cite{Pal:2015p332}. Further, MED was roughly assumed to be 3 times MFP which would roughly account for 95 percent of the total collected intensity.}
		\label{tab:table1}
	\end{center}
\end{table}

\clearpage

\noindent\textbf{Supplementary Note 12: Berry phase calculations.}
\begin{figure}[h]
	\centering{
		\includegraphics[scale=0.525]{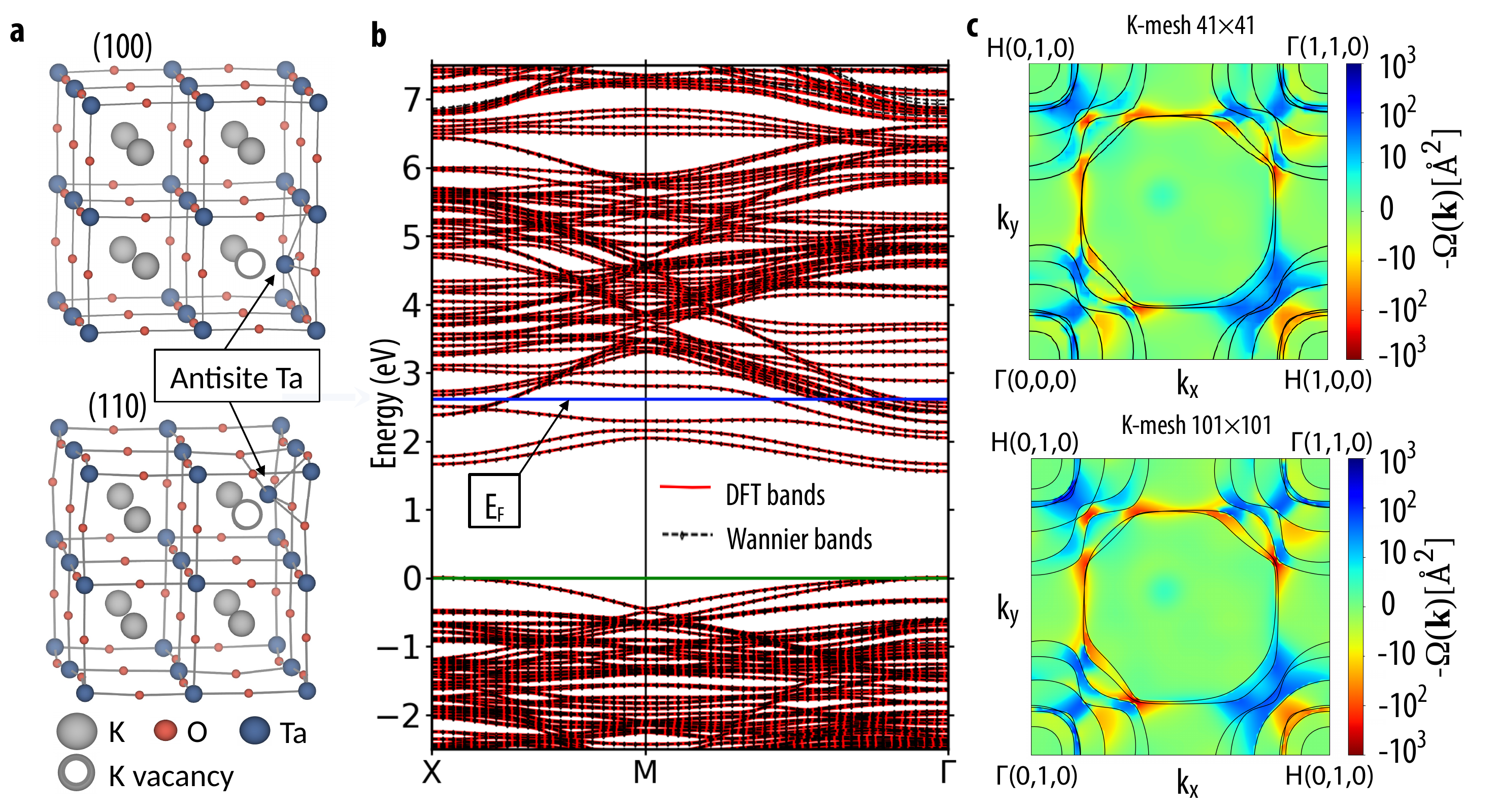}
		\caption{ \textbf{a.} The relaxed structure of KTaO$_3$ with Ta off-centering along the (100) and (110) direction around K vacancy (antisite like Ta defect) for a supercell of size 2$\cross$2$\cross$2. \textbf{b.} Band structure plot for (110) direction antisite like Ta defect calculated using density functional theory within PBE-GGA (shown in red solid lines) and Wannier interpolated bands (shown in black dotted lines). The blue solid line is the Fermi energy. \textbf{c.} Plot of total Berry curvature in the plane $k_z =0$ (in log scale) for the (110) antisitelike Ta defect structure calculated on a 2D k-mesh of sample-size $41 \cross 41$ (top panel) and $101 \cross 101$ (bottom panel). \label{fig:6}}
		\label{Fig6}}
\end{figure}

We have investigated the probable mechanism for the realization of PNRs theoretically using noncollinear density functional (DFT) theory. Antisite defect in perovskite materials like SrTiO$_3$ and complex perovskite oxides (Ca, Sr)$_3$Mn$_2$O$_7$ is known for causing macroscopic polarization in the system \cite{Jang:2010p197601, MiaoNCom4927:2022}. Due to the  similarities in the electronic properties of SrTiO$_3$ and KTaO$_3$, we expect KTaO$_3$ to develop polarization due to antsite defects. We have created Ta antisite defect in a supercell of size $2 \cross 2 \cross 2$ conventional unit cell. In our calculations, the Ta off-centering is considered along (100) and (110) directions. The relaxed structures are shown in Supplementary Fig. 12a. In Supplementary Fig. 12b we have shown the band structure of the system with Ta antisite defect along (110). The red solid lines are the bands calculated using DFT within PBE-GGA\cite{PBE:PRL3865:1996}. In order to calculate the macroscopic polarization in the system, we use the modern theory of polarization. The change in electronic contribution to the polarization $\Delta P$ is defined as \cite{VanderbiltPRB1651:1993, RestaRMP899:1994}

\begin{equation}
	\Delta P = -\dfrac{e}{2\pi}\phi  
\end{equation}
where $\phi$ is the Berry phase, which is the integral of the Berry curvature over a surface $S$ bounded by a closed path in $k-$space, \textit{i.e.,} \cite{BerryAMathPhys1802:1984}
\begin{equation}
	\phi = \int_S \Omega(\vb{k})d\vb{k}
\end{equation}
One can calculate Berry curvature using Bloch states $u_n(\vb{k})$ as
\begin{equation}
	\Omega_{\alpha \beta}(\vb{k})=\sum_n f_n(\vb{k}) \Omega_{n,\alpha\beta}(\vb{k}) = \sum_n -2 f_n(\vb{k})\text{Im} \braket{\pdv{u_n(\vb{k})}{k_\alpha}}{\pdv{u_n(\vb{k})}{k_\beta}} 
\end{equation}
where $\alpha, \beta$ are the cartesian indices and $f_n(\vb{k})$ is the occupation number of the state $n$. Ta antisite defects in KTaO$_3$ have a partially occupied band. As it well known, taking derivatives of  $u_n(\vb{k})$ in a finite-difference scheme in the presence of band crossing and avoided crossing becomes difficult. We followed the procedure described by X. Wang \textit{et. al.} \cite{VanderbiltPRB195118:2006} for the calculation of Berry curvature using  Wannier functions as implemented in the Wannier90 code \cite{wannier90JOP:2020}. In Supplementary Fig. 12b, we have shown the Wannier interpolated bands in black dotted lines. From the figure, it is clear that both the set of bands calculated using DFT and Wannier interpolation match very well near the Fermi level. In Supplementary Fig. 12c, we have shown the total Berry curvature for the antisite Ta defect along (110)  in the plane $k_z=0$ calculated using 2D k-mesh of size $41 \cross 41$ (top panel) and $101 \cross 101$ (bottom panel).  While there are small quantitative differences, the qualitative features in both the panels remain  the same.     
\clearpage

\clearpage

\noindent\textbf{Supplementary Note 13: Toy model of complex inter-band electronic relaxation mediated by a glassy bath.}

\vspace{0.5 cm}
As discussed in the main text, we consider the following model of coupled electron (el)-glass (gl) system.
\begin{subequations} \label{eq:Model_s}
	\begin{align}
		H&=H_{el}+H_{gl}+H_{el-gl}\\
		H_{el}&=-\sum_{i,j=1}^{N_c} t_{ij}c_i^\dagger c_j-\varepsilon_0\sum_{\alpha=1}^{N_f} f_\alpha^\dagger f_\alpha,~~~~(\varepsilon_0>0) \\
		H_{gl}&=\sum_{\mu=1}^{N_g} \frac{p_\mu^2}{2m}+U(\{x_\mu\})\\
		H_{el-gl}&=\sum_{i\alpha\mu} (V_{i\alpha\mu}c_i^\dagger f_\alpha+\mathrm{h.c.})x_\mu
	\end{align}
\end{subequations}
The electronic part consists of a conduction band and a flat impurity band at energy $-\varepsilon_0$. We consider three different energy dispersions for the conduction band, corresponding to different lattices and hopping amplitudes $t_{ij}$, namely -- (1) a flat band or $\delta$ function density of states (DOS) with bandwidth $W=0$, (2) a semicircular DOS, $g(\epsilon)=(1/2\pi)\sqrt{W^2-\omega^2}\theta(W-|\omega|)$ with band width $W$ [$\theta(x)$ is heaviside step function], e.g., corresponding to a  Bethe lattice, and (3) DOS corresponding to three-dimensional (3d) simple cubic lattice with nearest-neighbour hopping $W/12$. The band gap between the conduction band minimum and the impurity band is $\Delta=\varepsilon_0-W/2$. We set the chemical potential at the center of the conduction band, i.e., $\mu=0$, as appropriate for a metallic system (see Fig. 5(a) of main text. 

The Hamiltonian $H_{gl}$ for a set of particles with positions $\{x_\mu\}$, and their canonically conjugate momenta $p_\mu$ ($[x_\mu,p_\nu]=\imath \delta_{\mu\nu}$ with $\hbar=1$) models the dynamics of a local glassy background. The exact form of the inter-particle interaction $U(\{x_\mu\})$ is not crucial for our calculations. However, as an example, we take $U(\{x_\mu\})=\sum_{\mu<\nu<\gamma}J_{\mu\nu\gamma}x_\mu x_\nu x_\gamma$ with the spherical constraint $\sum_\mu x_\mu^2=N_g$, corresponding a well-known solvable model for glasses, namely the infinite-range spherical $p$-spin glass model with $p=3$-spin coupling \cite{Cugliandolo}. Here, $J_{\mu\nu\gamma}$ is real Gaussian random number with zero mean and variance $\overline{J^2_{\mu\nu\gamma}}=3!J^2/2N_g^2$, where the overline denotes averaging over different realizations of $J_{\mu\nu\gamma}$. The particular scaling with $N_g$ ensures extensive free energy in the thermodynamic limit $N_g\to \infty$. The $p$-spin glass model, with both quantum \cite{Cugliandolo}  and classical dissipative \cite{Sommers}  dynamics, undergoes a glass transition at temperature $T_g\sim J$. For temperature $T\gtrsim T_g$, the model gives rise to a supercooled liquid regime, like standard structural glasses \cite{Kob1997}, with complex two-step relaxation for the dynamical correlation function
\begin{align} \label{eq:GlassCorr_s}
	C_{gl}(t)&=\overline{\langle x_\mu(t)x_\mu(0)\rangle}=Ae^{-|t|/\tau_s}+Be^{-(|t|/\tau_\alpha)^\beta}
\end{align}
, having a short-time exponential decay, and long-time stretched exponential decay.

To obtain a solvable model for the coupled el-gl system, we also take the coupling $V_{i\alpha\mu}$ between the glass, and the conduction and impurity electronic states, infinite range. Here $V_{i\alpha\mu}$ is a complex Gaussian random number with zero mean and variance $\overline{|V_{i\alpha\mu}|^2}=V/(N_c N_f N_g)^{1/3}$. The particular scaling ensures a well-defined thermodynamic limit $N_c,N_f,N_g\to\infty$ with finite ratios $p_f=N_f/N_c$ and $p_g=N_g/N_c$. The ratios allow us tune the backaction of one part of the system on the other. For example, for simplicity, and to gain an analytical understanding, as discussed below, we take $p_f,p_g\gg 1$, such that the back actions of the conduction electrons on the impurity electrons and the glass are negligible. We expect our main conclusions to be valid for $p_f,p_g\approx 1$, i.e., when the backaction is substantial.

For the above-disordered model (Supplementary Eq. (10)), using the standard replica method \cite{Parisi,Cugliandolo}, we calculate the disorder-averaged connected equilibrium dynamical density-density correlation function for the conduction electrons, $C_{el}(t)=\overline{\langle n_i(t)n_i(0)\rangle}-\langle n_i(0)\rangle^2$, at temperature $T\ll \Delta$. Here $n_i=c_i^\dagger c_i$ is the conduction electron density. The correlation function $C_{el}(t)$ captures the relaxation of the thermally excited electrons in the conduction band at a finite temperature.

We obtain the replicated partition function, $Z^n=\int \mathcal{D}(\bar{c},c) \mathcal{D}(\bar{f},f)\mathcal{D}x \exp(-S[\bar{c},c,\bar{f},f,x])$, as coherent-state imaginary-time path integral over the fermionic Grassmann fields $\{\bar{c}_{ia},c_{ia},\bar{f}_{\alpha a},f_{\alpha a}\}$ and position variables $\{x_{ia}\}$, with replica index $a=1,\cdots,n$ and the action, 
\begin{align}
	S_n &= \int_0^{1/T} d\tau \sum_{ij, a}\cb_{ia}(\tau) [(\deltau-\mu)\delta_{ij}-t_{ij}] c_{ja}(\tau) +
	\int_0^{1/T} d\tau \sum_{\alpha, a} \fb_{\alpha a}(\tau) (\deltau-\mu-\varepsilon_0)f_{\alpha a}(\tau)
	\notag \\ 
	&+ \int_0^{1/T} d\tau \bigg[\frac{1}{2}\sum_{\mu,a}[m({\partial_\tau x_{\mu a}})^2+z x_{\mu a}^2] + \sum_{\mu\nu\gamma, a}J_{\mu\nu\gamma}x_{\mu a}(\tau)x_{\nu a}(\tau)x_{\gamma a}(\tau)\bigg]\nonumber \\ 
	&+\int_0^{1/T} d\tau \sum_{i\alpha\mu, a}[V_{i\alpha\mu} \cb_{ia}(\tau) f_{\alpha a}(\tau) x_{\mu a}(\tau) + \mathrm{h.c.}], 
\end{align}

where the Lagrange's multiplier $z(\tau)$ imposes the spherical constraint on $x_{\mu a}(\tau)$. After averaging over the distributions of $J_{\mu\nu\gamma}$ and $V_{i\alpha \mu}$, we obtain $\overline{Z^n}$ and the corresponding action as
\begin{align}
	\tilde{S}_{n} &= \int_0^{1/T} d\tau \sum_{ij, a}\cb_{ia}(\tau) [(\deltau-\mu)\delta_{ij}-t_{ij}] c_{ja}(\tau) +
	\int_0^{1/T} d\tau \sum_{\alpha, a} \fb_{\alpha a}(\tau) (\deltau-\mu-\varepsilon_0)f_{\alpha a}(\tau) \nonumber \\
	&+\frac{1}{2}\int^{1/T}_{0} d\t \sum_{\mu a}x_{\mu a}(\t)(-m\partial^2_{\tau}+z) x_{\mu a}(\t) +N_c\int d\t d\t'\sum_{ab} \bigg[V^2 (p_f p_g)^{1/3}G_{ab}(\t,\t')\mathcal{G}_{ba}(\tau',\tau) Q_{ab}(\t,\t')\nonumber \\
	&-\frac{J^2 p_g}{4} Q^3_{ab}(\t,\t')\bigg], 
\end{align}
where we have introduced the large $N$-fields,
\begin{subequations}
	\begin{align}
		G_{ab}(\t,\t')&=-\frac{1}{N_c}\sum_{i}c_{ia}(\t)\cb_{ib}(\t') \\
		\mathcal{G}_{ab}(\t,\t')&=-\frac{1}{N_f}\sum_{\alpha}f_{\alpha a}(\t)\fb_{\alpha b}(\t') \\
		Q_{ab}(\t,\t')&= \frac{1}{N_g}\sum_{\mu}x_{\mu a}(\t) x_{\mu b}(\t')
	\end{align}
\end{subequations}
To promote the above as fluctuating dynamical fields, conjugate fields $\Sigma_{ba}(\tau',\tau)$, $\sigma_{ba}(\tau',\tau)$ and $\Pi_{ab}(\tau,\tau')$ are introduced for $G,~\mathcal{G}$ and $Q$, respectively, e.g., by using the relation 
\begin{align}
	\int \mathcal{D}G \prod_{a\tau,b\tau'} \delta(N_c G_{ab}(\tau,\tau')+\sum_i c_{ia}(\tau)\bar{c}_{ib}(\tau'))=\int \mathcal{D}G\mathcal{D}\Sigma e^{-\int d\tau d\tau' \Sigma_{ba}(\tau',\tau)[N_c G_{ab}(\tau,\tau')+\sum_i c_{ia}(\tau)\bar{c}_{ib}(\tau')]}=1
\end{align}

As a result, we can now integrate out the fields $(\cb, c)$, $(\fb, f)$, and $x$. Assuming replica diagonal ansatz, e.g., $G_{ab}(\tau,\tau')=\delta_{ab}G(\tau,\tau')$, we obtain $\overline{Z^n}=\int \mathcal{D}(G,\mathcal{G},Q,\Sigma,\sigma,\Pi) e^{-nS_{eff}}$ and the effective action
\begin{align}
	S_{\rm eff} &= - N_c\int d\epsilon g(\epsilon) \Tr \ln \big(-\deltau+\mu-\epsilon-\Sigma \big)-N_f\Tr \ln\big(-\partial_\tau+\mu+\varepsilon_0-\sigma\big) +\frac{N_g}{2} \Tr \ln \big(-m\partial^2_{\t}+z -\Pi \big)\notag \\
	& -N_c\int d\t d\t' \big[\Sigma(\t,\t')G(\t',\t)+p_f \sigma(\t,\t')\mathcal{G}(\t',\t)-\frac{p_g}{2} \Pi(\t,\t')Q(\t',\t) +\frac{p_g J^2}{4} Q(\t,\t')^3\nonumber \\
	&-V^2 (p_fp_g)^{1/3}G(\tau,\tau')\mathcal{G}(\tau',\tau)Q(\tau,\tau')\big]  
\end{align}
In the large $N_c$ limit, the saddle point solution is obtained by varying the above action with respect to $G,\Sigma,\mathcal{G},\sigma,Q,\Pi$ and setting the variations to zero. Due to time-translation invariance at equilibrium, e.g., $G(\tau,\tau')=G(\tau-\tau')$ and $z(\tau)=z$. As a result, we can write the saddle point equations in the following form after performing a Matsubara Fourier transform first, e.g., $G(\tau-\tau')\to G(\imath \omega_n)$ with fermionic Matsubara frequency $\omega_n$, and then doing an analytical continuation, $G(\imath \omega_n)\to G_R(\omega+\imath 0^+$), to real frequencies, where $G_R$ is the retarded Green's function.
\begin{subequations} \label{eq:SaddleEq_s}
	\begin{align}
		G_R(\omega)&=\int d\epsilon g(\epsilon)\frac{1}{\omega+\mu-\epsilon-\Sigma_R(\omega)} \label{eq:GR_s}\\
		\mathcal{G}_R^{-1}(\omega)&=\omega+\mu+\varepsilon_0-\sigma_R(\omega)\\
		Q_R^{-1}(\omega)&=-m\omega^2+z-\Pi_R(\omega) \\
		\Sigma(\tau)&=V^2(p_fp_g)^{1/3}\mathcal{G}(\tau)Q(\tau)\\
		\sigma(\tau)&=V^2p_g^{1/3}p_f^{-2/3}G(\tau)Q(\tau)\\
		\Pi(\tau)&=\frac{3J^2}{2}Q^3(\tau)+V^2p_f^{1/3}p_g^{-2/3}G(\tau)\mathcal{G}(-\tau)
	\end{align}
\end{subequations}
In the limit $p_f,p_g\gg 1$, we can approximate $\sigma(\tau)\approx 0$ and $\Pi(\tau)\approx (3J^2/2)Q^3(\tau)$, and neglect the back action of the conduction electrons on the impurity states and the glass.

In this limit, following the numerical procedure similar to that in reference \cite{BanerjeeAltman, Bera}, we can numerically solve the above self-consistency equations for various conduction electron DOS $g(\epsilon)$ to obtain the retarded functions $G_R(\omega)$ and $Q_R(\w)$, whereas the retarded Green's function of the impurity band is given by $\mathcal{G}_R(\omega)=(\omega+\imath 0^++\mu+\varepsilon_0)^{-1}$. However, instead of self-consistently solving for $Q_R(\omega)$, we use the Supplementary Eq. (11) to obtain the spectral function of the glass from the fluctuation-dissipation relation
\begin{align}
	\rho_{gl}(\omega)&=-\frac{1}{\pi} \mathrm{Im}Q_R(\omega)=-\frac{1}{\pi} \tanh\left(\frac{\omega}{2T}\right)C_{gl}(\omega),
\end{align}
where $C_{gl}(\omega)=\int_{-\infty}^\infty dt e^{\imath \omega t}C_{gl}(t)$. As a result, the conduction electron self-energy can be obtained as
\begin{align} \label{eq:SigmaR_s}
	\Sigma_R(\omega) &= V^2 \int d\w_1 d\w_2\rho_f(\w_1)\rho_{gl}(\w_2) \frac{ n_F(\w_1)n_B(\w_2) - n_F(-\w_1)n_B(-\w_2)}{\omega-\omega_1-\omega_2+\imath 0^+},  
\end{align}
where $n_F(\omega)$ and $n_B(\omega)$ are Fermi and Bose functions, respectively, and $\rho_f(\omega)=\delta(\omega+\varepsilon_0)$ is the spectral function of the impurity electrons. Here we have redefined $V^2 (p_fp_g)^{1/3}$ as $V^2$. Thus, using $\Sigma_R(\omega)$ in Supplementary Eq. (17a)), we can obtain the conduction electron Green's function $G_R(\omega)$.

\subsection{Density density correlation function}
To capture the manifestation of the glassy relaxation in the electron-hole recombination process we look into the connected density-density correlator $C_{el}(t)=\langle n_i(t)n_i(0)\rangle -\langle n_i(0)\rangle^2$, which captures the relaxation of the thermally excited carriers at temperature $T$. $C_{el}(t)$ can be obtained from the imaginary-time  correlation function $C_{el}(\tau) = \langle n(\tau) n(0) \rangle-\langle n(0)\rangle^2$, where $n(\tau) = (1/N_c)\sum_{i}\cd_i(\tau)c(\tau)$. In the large-$N_c$ limit, the connected correlator is given by the bubble diagram and can be expressed as $C_{el}(\tau) = G(\tau)G(-\tau)$. Performing Matsubara Fourier transformation and then analytically continuing $i\Omega_n \to \omega + i0^+$, where $\Omega_m$ is bosonic Matsubara frequency, we can obtain the retarded correlator
\begin{align} \label{eq:CelR_s}
	C_{el,R}(\w) = \int d\w_1 d\w_2 \rho_c(\w_1) \rho_c(\w_2)\bigg(\frac{n_F(\w_1)n_F(-\w_2)-n_F(-\w_1)n_F(\w_2)}{\w_1-\w_2-\w_n-i0^+ } \bigg),
\end{align}
where $\rho_c(\omega)=-(1/\pi)\mathrm{Im}G_R(\omega)$ is the conduction electron spectral function. Using the fluctuation-dissipation theorem, we can relate $C_{el,R}(\omega)$ to the Fourier transform of the real-time density-density correlation function $C_{el}(t)$ as $C_{el}(\omega) = \coth(\omega/2T) \text{Im}C_{el,R}(\omega)$. Finally, performing the inverse Fourier transform we obtain the real-time density-density correlation function $C_{el}(t) = \int^{\infty}_{-\infty} (d\omega/2\pi) e^{-i\omega t}C_{el}(\omega)$. We show the results for the numerically computed $C_{el}(t)$ in Fig. 5(c) of the main text for (1) a flat conduction band ($W=0$), and (2) a semicircular conduction band DOS, using $C_{gl}(t)$ from Supplementary Eq. (11) (Fig. 5(b), main text). For the latter, we take a temperature-independent stretching exponent $\beta=0.5$ and the $\alpha$-relaxation time $\tau_\alpha(T)$, which varies as $\sim T^{-2.8}$, consistent with our experimental results (Fig. 2(c), main text). We also vary the coefficient $B(T)\leq 1$ such that it increases with decreasing temperature, while $A(T)=1-B(T)$. This leads to a more dominant long-time stretched exponential part in $C_{gl}(t)$ at lower temperatures, compared to the short-time exponential decay in Supplementary Eq. (11). In Supplementary Fig. 13, we show the results for the conduction band energy dispersion corresponding to a nearest-neighbour tight binding model on a simple cubic lattice. We can see that in all the cases the glassy relaxation of the bath influences the relaxation of the conduction electrons leading complex and temperature dependent relaxation profile for $C_{el}(t)$.

We can obtain a simple analytical understanding of the above results as  follows. From Supplementary Eq. (19), we can obtain for $\mu=0$

\begin{figure}[h]
	\centering{
		\includegraphics[scale=0.8]{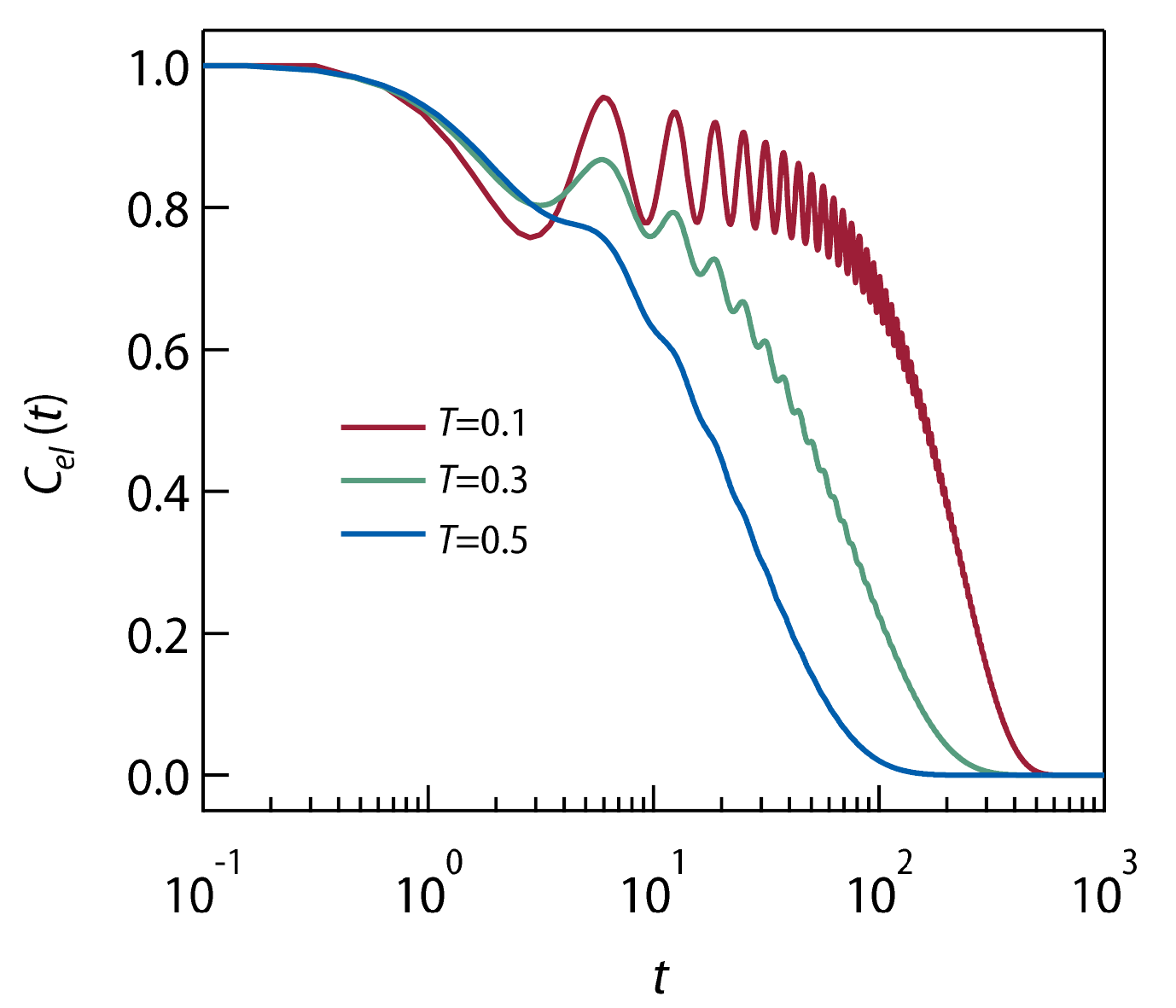}
		\caption{ The density-density correlation function of conduction electron $C_{el}(t)$ vs. $t$ for three temperatures corresponding to a nearest-neighbour tight binding model on a simple cubic lattice }
		\label{Fig9}}
\end{figure}

\begin{align}
	\gamma(\w) = {\rm Im }\Sigma_R(\omega) &= -\pi V^2  \rho_{gl}(\w +\varepsilon_0) \big( n_F(-\varepsilon_0) + n_B(-\varepsilon_0-\w) \big)  
\end{align}

Thus, by defining $k(\omega)=\mathrm{Re}\Sigma_R(\omega)$, we obtain from Supplementary Eq. (17a)

\begin{align}
	{\rm Im }G_R(\w) &= \int d\epsilon g(\epsilon) \frac{\gamma(\w)}{(\w-\epsilon-k(\w))^2+\gamma(\w)^2}
\end{align}
 
As a result, from Supplementary Eq. (20) we get
\begin{align}
	{\rm Im}C_{el,R}(\w) &= \int d\w_1 \rho_c(\w_1) \rho_c(\w_1-\w)\big(n_F(\w_1)-n_F(\w_1-\w)\big)
\end{align}

The above can be expressed as 
\begin{align}
	{\rm Im}C_R(\w)  &= \frac{1}{\pi^2}\int  d\epsilon_1 d\epsilon_2 g(\epsilon_1)g(\epsilon_2) \int d\w_1 \bigg[\frac{\gamma(\w_1)}{(\w_1-\epsilon_1-k(\w_1))^2+\gamma(\w_1)^2} \nonumber \\ &\frac{\gamma(\w_1-\w)}{(\w_1-\w-\epsilon_2-k(\w_1-\w))^2+\gamma(\w_1-\w)^2}\bigg] \big(n_F(\w_1)-n_F(\w_1-\w)\big)
\end{align}
Thus,
\begin{align}
	C_{el}(t)=\int_{-\infty}^\infty d\omega_1d\omega e^{-\imath \omega t}\rho_{gl}(\omega_1+\varepsilon_0)\rho_{gl}(\omega_1+\varepsilon_0-\omega) F(\omega,\omega_1)
\end{align}
can be written as a convolution over the glass spectral function $\rho_{gl}(\omega)$, where
\begin{align}
	F(\omega,\omega_1)&=\frac{V^4}{2\pi}\int d\epsilon_1 d\epsilon_2 \frac{g(\epsilon_1)g(\epsilon_2)\coth\left(\omega/2T\right)[n_F(\w_1)-n_F(\w_1-\w)]}{[(\w_1-\epsilon_1-k(\w_1))^2+\gamma(\w_1)^2][(\w_1-\w-\epsilon_2-k(\w_1-\w))^2+\gamma(\w_1-\w)^2]}\nonumber \\
	& \big( n_F(-\varepsilon_0) + n_B(-\varepsilon_0-\w_1) \big) \big( n_F(-\varepsilon_0) + n_B(-\varepsilon_0-\w_1+\omega) \big)
\end{align}
The spectral function of the glass contains the information about the multiple time scales and the non-trivial temperature dependence of the glassy relaxation. Thus if appropriate conditions on the electronic energy scales $W$ and $\varepsilon_0$ ($\Delta$) are made relative to the energy scales of the glass, e.g., $\tau_s^{-1}$ and $\tau_\alpha^{-1}$, then the complex relaxation of the conduction electrons can be obtained. We numerically find that these conditions are met if the glassy bath is broad, i.e., the bandwidth of the glass is comparable or larger than the electronic energy scales.

\end{document}